\documentclass[twocolumn,aps,prb]{revtex4-2}

\usepackage{graphicx}
\usepackage[dvipsnames]{xcolor}

\definecolor{bdiv2-0}{HTML}{7ed9e1}
\definecolor{bdiv2-1}{HTML}{58aebf}
\definecolor{bdiv2-2}{HTML}{3f8597}
\definecolor{bdiv2-3}{HTML}{345a68}
\definecolor{bdiv2-4}{HTML}{2b343e}
\definecolor{bdiv2-5}{HTML}{4f2330}
\definecolor{bdiv2-6}{HTML}{91263c}
\definecolor{bdiv2-7}{HTML}{bd494c}
\definecolor{bdiv2-8}{HTML}{de725a}
\definecolor{bdiv2-9}{HTML}{ff9b68}

\usepackage[
colorlinks=true,
linkcolor=bdiv2-6,
citecolor=bdiv2-3,
urlcolor=bdiv2-2,
]{hyperref}

\usepackage{xspace}
\usepackage{amsmath}
\usepackage{amsfonts}
\usepackage{amssymb}
\usepackage{mathtools}
\usepackage{braket}
\usepackage{comment}
\usepackage[capitalize]{cleveref}
\usepackage{sidecap}

\newcommand{\vdagger}{\vphantom{\dagger}}
\newcommand{\cre}{c^{\dag}}
\newcommand{\ann}{c^{\vphantom{\dag}}}
\newcommand{\crespinor}{\vec{c}^{\, \dag}}
\newcommand{\annspinor}{\vec{c}^{\vphantom{\dag}}}
\newcommand{\I}{\mathrm i}
\newcommand{\sym}{\mathcal S}
\newcommand{\ph}{\mathcal P}
\newcommand{\ex}{\mathcal E}
\newcommand{\chiral}{\mathcal C}
\newcommand{\TRS}{\mathcal T}
\newcommand{\e}{\mathrm e}
\newcommand{\dd}{\mathrm d}
\newcommand{\bvec}[1]{\boldsymbol{#1}}
\newcommand{\Ef}{E_\text{F}}

\usepackage{tikz}
\colorlet{c-v-pppp}{bdiv2-1!70!bdiv2-0}
\colorlet{c-v-pmpm}{bdiv2-4!10!bdiv2-3}
\colorlet{c-v-pmmp}{bdiv2-5!10!bdiv2-6}
\colorlet{c-v-ppmm}{bdiv2-8!70!bdiv2-9}
\definecolor{diffgray}{RGB}{180,180,180}

\usetikzlibrary{arrows, positioning, calc, shapes, decorations.markings,
decorations.pathreplacing, math, snakes, arrows.meta, angles, hobby}
\tikzset{pppp/.style={decorate,decoration={snake,amplitude=0.6mm,segment
length=2.3mm,pre length=0mm,post length=0mm},c-v-pppp, thick}}
\tikzset{pmpm/.style={decorate,decoration={coil,amplitude=1mm,segment
length=1.6mm,pre length=0mm,post length=0mm},c-v-pmpm, thick}}
\tikzset{pmmp/.style={dashed, c-v-pmmp, very thick}}
\tikzset{ppmm/.style={decorate,decoration={zigzag,amplitude=0.6mm,segment
length=3.3mm,pre length=0mm,post length=0mm},c-v-ppmm, thick}}
\tikzset{mf/.style={double distance = .2mm, diffgray, thick}}
\tikzset{midarrow/.style={decoration={markings,mark=at position 0.5 with
{\arrow[xshift=2.5pt]{Latex[length=4pt,#1]}}},postaction={decorate}}}
\tikzset{vp/.style={midway,outer sep=2.5pt,node contents={\tiny$+$}}}
\tikzset{vm/.style={midway,outer sep=2.5pt,node contents={\tiny$-$}}}

\usepackage{makerobust}
\newcommand{\makeauthor}[2]{\newcommand{#1}[1]{{%
  \protect%
  \color{#2}{%
    \bfseries%
    \begingroup\escapechar=-1\edef\x{\endgroup\string#1}\x:%
  }\itshape{} ##1}}%
  \MakeRobustCommand#1}
\makeauthor{\md}{Plum}
\makeauthor{\lk}{ForestGreen}
\makeauthor{\mk}{orange}

\newcommand{\sAM}[0]{$s$AM\xspace}
\newcommand{\sAMs}[0]{$s$AMs\xspace}
\newcommand{\prlparagraph}[1]{\textit{#1}---}

\newcommand{\supplement}[1]{%
  \clearpage%
  \title{#1}%
  \maketitle%
  \setcounter{equation}{0}%
  \setcounter{figure}{0}%
  \setcounter{table}{0}%
  \setcounter{page}{1}%
  \makeatletter%
  \renewcommand{\thesection}{S\arabic{section}}%
  \renewcommand{\thesubsection}{\Alph{subsection}}%
  \renewcommand{\theequation}{S\arabic{equation}}%
  \renewcommand{\thefigure}{S\arabic{figure}}%
  \renewcommand{\thetable}{S\Roman{table}}%
  \renewcommand{\thepage}{S\arabic{page}}%
  \numberwithin{figure}{section}%
  \numberwithin{table}{section}%
  \numberwithin{equation}{section}%
  \makeatother%
  \onecolumngrid%
}
\makeatletter
\def\maketitle{
\@author@finish
\title@column\titleblock@produce
\suppressfloats[t]}
\makeatother

\let\oldcite\cite
\renewcommand{\cite}[1]{\if\relax\detokenize{#1}\relax\textbf{\color{red}[?]}\else\oldcite{#1}\fi}

\begin{document}
\title{Extended {\itshape s}-wave altermagnets}
\author{Matteo Dürrnagel}
\thanks{These authors contributed equally.}
\author{Lennart Klebl}
\thanks{These authors contributed equally.}
\author{Tobias Müller}
\author{Ronny Thomale}
\author{Michael Klett}
\email{michael.klett@uni-wuerzburg.de}
\affiliation{Institut für Theoretische Physik und Astrophysik and
Würzburg-Dresden Cluster of Excellence ctd.qmat, Universität Würzburg, 97074
Würzburg, Germany}

\date{\today}

\begin{abstract}
We propose extended $s$-wave altermagnets (\sAMs) as a class of magnetic states which are fully gapped, spin-compensated, and feature spin-polarized bands.
\sAMs are formed through valley-exchange symmetries, which act as momentum-space translations beyond standard crystallographic spin-group classifications. Using an effective two-valley model, we demonstrate that \sAMs exhibit isotropic spin splitting, enable spin-selective transport in tailored heterostructures, and give rise to descendant pair density wave order. From a microscopic \sAM minimal model, we develop the guiding principles to identify \sAMs in quantum magnets.

\end{abstract}

\maketitle

\prlparagraph{Introduction}%
Altermagnetism~\cite{Smejkal2020, Naka2019, Ahn2019, Hayami2019, Yuan2020, Samanta2020c} has generated tremendous interest in condensed matter physics~\cite{Smejkal2022, Smejkal2022a, Smejkal2022b, Cheong2024, Roig2024m, Tamang2024, Song2025a, Rodrigo2025a, Song2025e, Remi2025a, Jungwirth2025a}. Distinct from ferromagnets and antiferromagnets, altermagnets exhibit a wealth of magneto-electric phenomena while retaining vanishing net magnetization, rendering them promising candidates for spintronic technologies~\cite{Gonzalez2021, Shao2021, Smejkal2022, Guo2024, Zeng2024, jungwirth2025altermagnetic,Sicheler2025o}. It originates from spin-split electronic bands in certain regions of the Brillouin zone, where magnetic compensation is enforced by crystallographic space-group symmetries. In the case of altermagnets, the magnetic order parameter transforms according to a nontrivial irreducible representation of magnetic space groups, leading to nodal structures on the Fermi surface of $p$-, $d$-, $f$-, $g$-wave type or higher~\cite{Smejkal2022a, Roig2024m}. The nodes may underpin unconventional transport and topological responses, but they also pose challenges for device applications. In particular, low-energy excitations associated with the nodes induce quasiparticle poisoning, limiting both stability and performance.   

Comparing the magnetic translation group profile against non-magnetic space groups allows to draw certain connections to the spectral nature of superconductors (SC) and their associated point group symmetry.
For an SC condensate, the Bogoliubov spectrum is governed by the single-particle Fermiology in the Brillouin zone (BZ), together with the irreducible point group representation of the SC, which encodes the relative angular momentum $l$ of the electrons forming a Cooper pair. 
Finite $l$ typically implies a nodal line pattern within the BZ, which generically intersects with the Fermi surface and leads to nodes. 
For $l=0$, the lowest lattice harmonic associated with the Cooper pair wave function is a constant and thus, without any sign change, yields a full gap. 
This case is usually referred to as conventional $s$-wave superconductivity, as its microscopic mechanism typically arises from electron–phonon coupling. In contrast, a sign-changing Cooper pair wave function results from unconventional pairing mediated by repulsive electronic interactions. 
However, as first conceived for the iron pnictides~\cite{PhysRevLett.101.057003,PhysRevB.78.134512,MAZIN2009614}, a multi-pocket Fermi surface allows for an unconventional $l=0$ SC pairing mechanism where different Fermi pockets can feature a sign-changing SC order parameter yet no nodes, as these do not intersect the Fermi pockets centered around $\Gamma$ and M.

\begin{figure}
    \centering
    \includegraphics{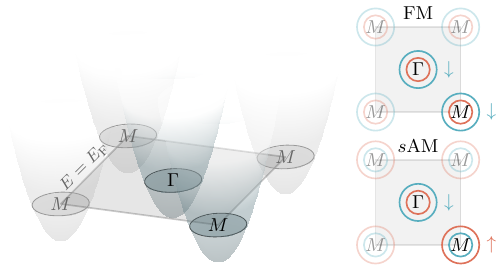}
    \caption{Sketch of an $s$-wave altermagnetic two-valley system. 
    Left: schematic single-particle spectrum of the two valleys ($\nu=\Gamma,M$) described by \cref{eqn:H0_valley}. 
    Right: spin-split Fermi surfaces for the ferromagnetic (top) and \sAM (bottom) order parameters, 
    $\Delta^\text{FM}\tau^0\sigma^z$ and $\Delta^\text{\sAM}\tau^z\sigma^z$, respectively. 
    In the \sAM case, valley symmetry enforces magnetic compensation through $\tau^z$, 
    so the spin-up (red, $\color{bdiv2-7}\uparrow$) and spin-down (blue, $\color{bdiv2-2}\downarrow$) Fermi surfaces remain equal in size.}
    \label{fig:valleys}
\end{figure}

In this Letter, we propose a new class of effectively gapped, spin-compensated magnets with spin-polarized bands, which we call extended $s$-wave altermagnets (\sAMs). A construction of such magnetic states has so far been absent from the taxonomy of altermagnets~\cite{Cheong2024, Roig2024m, Shao2025c}. By analogy, as much as the finite $l$ particle-particle SC condensate relates to the finite $l$ altermagnetic particle-hole condensate, the conventional $l=0$ SC relates to a ferromagnet with, in contrast to antiferromagnets and altermagnets, uncompensated total magnetic flux. The key aspect to appreciate now is that, in full correspondence to unconventional SC defined by the presence or absence of momentum dependence of Cooper pair wave function, (un)conventional magnetic particle-hole pairs can be defined by the presence (absence) of momentum depdence of the particle-hole pair wave function. Following this rational, ferromagnets and antiferromagnets are conventional, while altermagnets are unconventional. Furthermore, similar to the distinction between $s$-wave and extended $s$-wave SC, the specification of the magnetic space group to contain (anti-)ferromagnets alone does not sufficiently characterize the state to be a conventional magnet. Rather, it is the harmonic composition, i.e., constant vs. $k$-dependent harmonics, that decides about the given magnetic state to be conventional or unconventional. The  \sAM state we introduce in this work bears such similarity to an unconventional extended $s$-wave SC pairing, as its momentum-dependent particle-hole wave function features sign changes between different Fermi pockets but no nodes, as its Umklapp surface nodes do not intersect with the Fermi pockets. 

We formulate \sAMs as staggered spin-polarized states in a multi-valley continuum model, where momentum-space translation between valleys enforces magnetic compensation, and subsequently present a minimal lattice model that realizes the \sAM state. As we explore their transport signatures, we find that \sAMs enable spin-selective current control, and as such provide an ideal platform for spintronics state engineering tentatively superior to previous altermagnetic proposals.

\prlparagraph{Valley model}%
\Cref{fig:valleys} depicts a two-valley system ($\nu=\Gamma,M$) on a
two-dimensional square lattice. One isotropic electron pocket lies at the zone
center $\nu=\Gamma$, while the second is located at the zone corner $\nu=M$. Both pockets
share the same quadratic dispersion, leading to the non-interacting continuum
Hamiltonian
\begin{equation}
    H_c = \int_{|\bvec k|<k_c} \!\!\dd\bvec k~
    \sum_{\nu,\sigma} \frac{{\bvec k}^2}{2m}\,
    c^\dagger_{\bvec k+\nu,\sigma}\, c^{\vdagger}_{\bvec k+\nu,\sigma} \,,
    \label{eqn:H0_valley}
\end{equation}
where $c^{(\dagger)}_{\bvec k+\nu,\sigma}$ annihilates (creates) an electron of mass $m$ with spin $\sigma$ in valley $\nu$. The momentum cutoff $k_c$ ensures the
validity of the valley description within a periodic lattice. \Cref{eqn:H0_valley} is symmetric under rotations in combined spin-valley space.

In the presence of electron-electron interactions, Coulomb repulsion is expected to discriminate between intra- and inter-valley scattering processes corresponding to momentum transfers $\bvec q \approx 0$ and $\bvec q \approx M$, respectively. As a result, the spin-valley $SU(4)$ symmetry of the non-interacting Hamiltonian limit is reduced to a direct product of valley-$U(1)$~\footnote{If inter-valley Hund's coupling is present, valley-$U(1)$ is expected. For density-density type inter-valley interactions, the Hamiltonian remains symmetric under arbitrary rotations in valley space resulting in valley-$SU(2)$~\cite{Fischer2024s}.} and spin-$SU(2)$. This lower symmetry constrains the set of allowed magnetic orders to three distinct classes~\cite{SM}: spin polarization (SP, $\tau^0\sigma^z$), staggered spin polarization (SSP, $\tau^z\sigma^z$), and spin inter-valley coherence (SIVC, $\tau^{x,y}\sigma^z$). Here, $\tau^i$ and $\sigma^i$ denote Pauli matrices acting in valley and spin space, respectively and $\tau^x$ (i.e., valley exchange) presents a symmetry of the non-magnetic Hamiltonian.

Notably, these symmetry structures have recently been connected to momentum-space nonsymmorphic space groups, in particular to the group $P4g$. In this context, our models' valley exchange operations can be understood naturally as symmetry-based fractional momentum space translations~\cite{Zhang2023g}.

\begin{figure}
    \centering
    \includegraphics{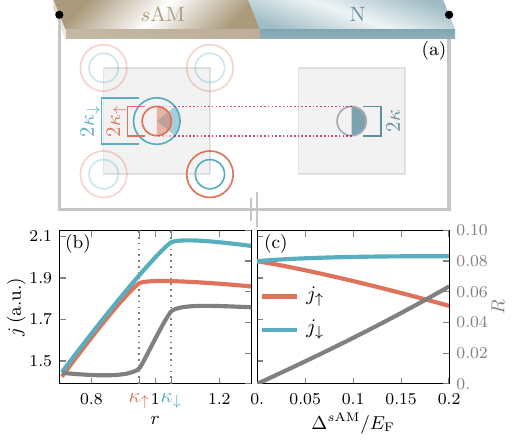}
    \caption{Spin splitter effect in an \sAM{}-N junction. (a)~Setup of the \sAM-N
    heterostructure with the Fermi surfaces and their respective radii $\kappa_{(\uparrow/\downarrow)}$ indicated. Due to the reduced modes
    available on the N side of the junction, only electronic eigenstates in the
    shaded areas can propagate through the interface and contribute to the
    current. Panels~(b,c) display the spin polarized currents $j_\sigma$ (red, blue) and the spin conversion factor $R=|(j_\uparrow-j_\downarrow)/(j_\uparrow+j_\downarrow)|$ (gray) as a function of (b)~Fermi surface mismatch $r = 2\kappa / (\kappa_\uparrow+\kappa_\downarrow)$ and (c)~\sAM order parameter $\Delta^\text{\sAM}$ in units of the Fermi energy $E_\mathrm{F}$. See Supplemental Material~\cite{SM} for simulation parameters.}
    \label{fig:junction}
\end{figure}

At mean-field level and one-loop renormalization group level the system is found to favor magnetic order in the SSP channel~\cite{SM}. The SSP order parameter ($\tau^z\sigma^z$) breaks both valley and spin
degeneracy, resulting in oppositely spin-split bands in the two valleys
(cf.~\cref{fig:valleys}). 
Notably, the SSP state breaks the combined parity--time ($\mathcal{PT}$) symmetry, which would otherwise enforce spin degeneracy at each fixed $ \bvec k $; this $\mathcal{PT}$-breaking is precisely what allows spin-split bands despite magnetic compensation, i.e., a characteristic altermagnetic fingerprint.
The system is still invariant under an
exchange of valleys and a subsequent spin flip, i.e., $\big[ H_c + \, \Delta^\text{SSP}, \tau^x\sigma^x \big] = 0 $ with 
\begin{equation}
    \Delta^\text{SSP} \propto \int_{|\bvec k|<k_c} \!\!\dd\bvec k~ \sum_{\sigma, \nu} \cre_{\bvec k + \nu, \sigma} \tau^z_{\nu\nu} \sigma_{\sigma\sigma}^z  \ann_{\bvec k + \nu, \sigma} \,.
\end{equation}
Valley exchange $\tau^x$ thus acts as a symmetry that allows for magnetic
compensation. The extended $s$-wave character is rooted in the position of one
pocket at the zone center $\nu=\Gamma$ and the other at $\nu=M$. Importantly, $\tau^x$ does not
necessarily resemble a simple real-space symmetry operation, as it corresponds
to a shift of $\bvec q=M$ in momentum space. This amounts to generalizing the classification of altermagnetism to symmetries of other origin than the crystallographic space-group naturally, and leads us to the identification of the SSP state as an \sAM: A
spin-compensated (collinear) magnet with spin-split bands and a nodeless gap
structure, cf.~\cref{fig:valleys}.
It is important to emphasize that in most multi-valley material platforms, valley exchange appears only as an emergent low-energy symmetry of the electron gas~\cite{Schaibley2016v}. Consequently, the symmetry-driven spin compensation of the \sAM state is strictly valid only within the energy range where the electronic bands remain fully valley-polarized. For low-energy transport phenomena, however, as sought after in spintronics applications, it is precisely these states dominating the system’s response.

\prlparagraph{Spin-polarized transport}%
In conventional altermagnets, direction-dependent spin filtering arises naturally from nodal structures in the quasiparticle spectrum~\cite{Smejkal2022}. The situation in the \sAM state is more subtle. Here, the central exploitable feature for spin-selective transport is the radial separation of spin-polarized valleys in momentum space ($\nu=\Gamma,M$). \Cref{fig:junction}(a) illustrates a possible device geometry: An \sAM (gold, left) is coupled to a normal semi-metallic conductor (N, metallic blue, right) to form an \sAM{}-N heterojunction.
Because of the isotropic spin splitting and the valley symmetry of the \sAM state, transport averaging over the full BZ yields a cancellation of spin polarization. Momentum-space filtering of states that participate in transport~\cite{kashiwaya1996theory, cayssol2008crossed, linder2008tunneling, breunig2018creation, Breunig2021d}, however, lifts this compensation by the mismatch of available states on either side on the \sAM{}-N interface.
In particular, the normal conductor sketched in \Cref{fig:junction}(a) provides propagating modes only near the zone center, while modes near the zone corner are absent. As a result, only the spin-polarized $\Gamma$ pocket contributes to transmission across the interface, effectively turning the junction into a spin splitter. The spin conversion efficiency is quantified by the spin conversion factor $R = \left|\frac{j_\uparrow - j_\downarrow}{j_\uparrow + j_\downarrow}\right|$,
where $j_\sigma$ denotes the spin polarized current~\cite{SM}.
Depending on model parameters, we obtain $R$ values comparable to those of nodal altermagnets (see \cref{fig:junction}(b,c) and, e.g., Refs.~\cite{Zelezny2017s, Naka2019s, Naka2021p, Giuli2024, Lai2025d}). Unlike the nodal altermagnetic case, however, the spin splitting for the \sAM is isotropic, and can be tuned by the choice of normal conductor (N). The setup depicted in \cref{fig:junction}(a) acts like a conventional (single-pocket) ferromagnet in transport. Using a normal conductor with an electron pocket around $M$ instead of $\Gamma$ effectively reverses the spin orientation, which opens the door for the construction of mesoscopic spin-splitter devices. Furthermore, the \sAM's resilience against disorder will be enhanced due to the homogeneous excitation gap. 
The design sketched in \cref{fig:junction} serves as a proof of principle for the spintronic device application of \sAM heterostructures using spin neutral currents. In close analogy to nodal AMs, we expect the unique spin splitting of \sAMs to transpire as a versatile tool in other heterostructure arrangements, too~\cite{Shao2021, Sun2025t}, exploiting, e.g., tunneling magnetoresistance to generate spin selective transport across an \sAM{}-insulator-FM junction.

\prlparagraph{Superconductivity}Beyond transport, altermagnets are known to host exotic superconducting descendant phases due to spin-polarized Fermi surfaces~\cite{Sim2025p,Peng2025v,Parthenios2025spin}. For our \sAM, spin polarization of each valley suppresses ordinary $\bvec q=0$ spin-singlet pairing leaving only two possible channels: (i)~intra-pocket spin triplet pairing and (ii)~inter-pocket spin singlet pairing. For~(i), electrons from the same valley can form spin-triplet Cooper pairs, necessitating odd parity in real space. This implies $p$-wave (or higher) gap nodes, tending to reduce the condensation energy and rendering the state energetically unfavorable. In case of inter-pocket spin-singlet pairing~(ii), electrons from opposite valleys can pair with total momentum $\bvec q = M$. Thanks to the residual symmetry of the \sAM state, the entire Fermi surface can participate, stabilizing inter-pocket spin singlet pairs with finite momentum. This mechanism is thus expected to promote a pair density wave state~\cite{annurev:/content/journals/10.1146/annurev-conmatphys-031119-050711}, which remains fully gapped. It is therefore expected to be more robust than the intrinsically nodal superconducting orders that emerge from $d$-wave altermagnetic parent states~\cite{Sim2025p, Peng2025v}.

\begin{figure}
    \centering
    \includegraphics{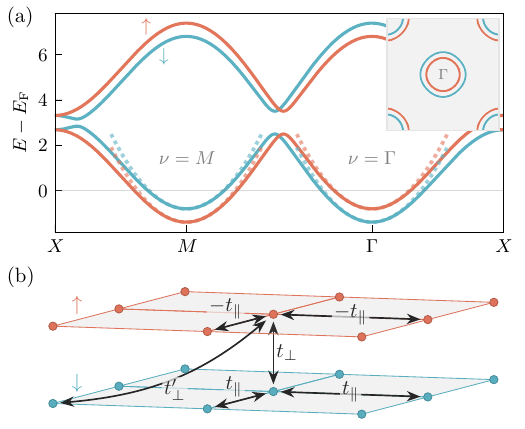}
    \caption{$s$-wave altermagnetic lattice model. Panel~(a) displays the altermagnetic band structure along the irreducible path, with spin-up/down indicated by red/blue, respectively. The dotted lines demonstrate how the lattice model \cref{latticemodel} is approximated by the two-valley model \cref{eqn:H0_valley}. The top-right inset shows the Fermi surface in the Brillouin zone. Panel~(b) illustrates the real-space hopping processes that lead to the band structure in panel~(a). We set $t_\parallel=1$, $t_\perp=0.5$, $t_\perp^\prime=0.1$, and $\Delta^\text{\sAM} = 0.3$. The chemical potential is close to the band bottom of the nonmagnetic system, i.e., $E_\mathrm{F}=-3$.}
    \label{fig:bandplot}
\end{figure}

\prlparagraph{Lattice model}%
Consider the bilayer hopping model
\begin{multline}
    H_l = -t_\parallel \sum_{\braket{ij},\sigma} {\vec c}^{\,\dagger}_{i,\sigma} \tau^z {\vec c}^{\vdagger}_{j,\sigma} - t^\prime_\perp \sum_{\braket{\braket{ij}},\sigma} {\vec c}^{\,\dagger}_{i,\sigma} \tau^x {\vec c}^{\vdagger}_{j,\sigma} \\
    - t_\perp\sum_{i,\sigma} {\vec c}^{\,\dagger}_{i,\sigma} \tau^x {\vec c}^{\vdagger}_{i,\sigma} \,,\label{latticemodel}
\end{multline}
as sketched in \cref{fig:bandplot}(b). Here, ${\vec c}^{\,(\dagger)}_{i,\sigma}$ denotes the two-component annihilation (creation) operator in layer space with $\tau^i$ denoting Pauli matrices spanning this space, while $\braket{ij}$ and $\braket{\braket{ij}}$ indicate sums over nearest and next-nearest neighbor bonds, respectively.
In momentum space \cref{latticemodel} reads $H_l = \sum_{\bvec k,\sigma} {\vec c}^{\,\dagger}_{\bvec k, \sigma}
    \big[v_x(\bvec k) \tau^x+ v_z(\bvec k) \tau^z\big]
    {\vec c}^{\vdagger}_{\bvec k, \sigma} \,,$
with the in-plane and out-of-plane dispersions
\begin{equation}
\begin{aligned}
    v_x(\bvec k) &{}= -t_\perp-4t_\perp^\prime\cos k_x\cos k_y \,, \\
    v_z(\bvec k) &{}= -2t_\parallel(\cos k_x + \cos k_y) \,.
\end{aligned}
\end{equation}
\Cref{latticemodel} possesses $C_{4v}$ symmetry and is further invariant under $\mathcal S {\vec c}^{\,\dagger}_{\bvec k, \sigma} \mathcal S^{-1} = \tau^x {\vec c}^{\,\dagger}_{\bvec k+M, \sigma} $,
i.e., a momentum shift by $\bvec q=M$ combined with a layer exchange. As pointed out above, this can be viewed as a lattice realization of the continuum valley-exchange symmetry This corresponds to an inverted layer polarization of the Bloch functions upon shifting the Brillouin zone by $\bvec q=M$, while retaining an identical dispersion.

The \sAM state in \cref{latticemodel} is realized through a staggered ferromagnetic order of the form $\Delta^\text{\sAM} \propto \tau^z\sigma^z$. Similar to the valley model in \cref{eqn:H0_valley}, a spin flip can be generated through $\mathcal S$, i.e., the magnetic Hamiltonian $H_l + \Delta^\text{\sAM}$ is invariant under the symmetry operation $\{\mathcal{S} \parallel C_2\}$, where the first element acts in momentum/orbital and the second in spin space. Notably, since $\mathcal S \notin C_{4v}$, this symmetry is not an element of the spin group associated with the crystallographic space group.
The \sAM state is therefore protected by $\mathcal S$. At the same time, it is fully invariant under all space-group operations of $C_{4v}$, which directly implies a spin gap of (extended) $s$-wave symmetry. Indeed transforming the orbital structure of the magnetic gap to band space reveals a momentum dependent spin splitting with form factor 
\begin{equation}
    \Delta^\text{\sAM}(\bvec k)\propto \big(\cos k_x+\cos k_y\big) + \mathcal{O}\big((\cos k_x+\cos k_y)^2\big)\,,
\end{equation} in full analogy to the momentum-dependent $l\neq 0$ gap structure of altermagnets. \Cref{fig:bandplot}(a) shows the band structure of the lattice model in the \sAM state. We find that the valley model \cref{eqn:H0_valley} is the correct low-energy effective theory of \cref{latticemodel} around the two valleys $\nu=\Gamma,M$, as indicated by the dotted lines. Moreover, the Fermi surface demonstrates the extended $s$-wave character of the magnetic state. Note that \cref{latticemodel} offers to include other couplings, e.g., $v_y(\bvec k)\tau^y$, and still retains its symmetries~\cite{SM}.

As the \sAM is protected by a symmetry $\mathcal S$ residing outside the point group $C_{4v}$, it is not only constrained to the irreducible representation (irrep) $A_1$, but even to specific lattice harmonics within $A_1$. The symmetry $\mathcal S$ pins the nodes of the \sAM gap to a certain region in momentum space (the $XY$-lines, see \cref{fig:bandplot}(a) and Supplemental Material~\cite{SM}). Hence only those lattice harmonics within the $A_1$ irrep that have nodes on the $XY$-lines are allowed in $v_z(\bvec k)$. Different gap structures---in terms of nodal lines---can be realized by changing the underlying symmetry $\mathcal S$~\cite{SM}.

\prlparagraph{Material candidates}\sAMs may be realized in any multi-valley system with one valley residing at $\nu=\Gamma$, regardless of the location of the other valley in momentum space---as long as valley exchange is implemented at least as an emergent effective symmetry at low energies. Beyond what is explicated above, this further allows to consider one-dimensional systems to feature \sAMs, as well as lattices that are not $C_{4v}$ symmetric~\cite{SM}. In addition to valley-based systems, we learn from \cref{latticemodel} which kind of momentum translation symmetry may be implemented to reach an \sAM (cf.~\cref{fig:bandplot}). Indeed, band structures with approximately correct momentum space symmetries have been observed in materials such as ZrSiS and ZrSiSe, though these materials are predicted to be excitonic insulators rather than displaying long range magnetic order~\cite{Rudenko2018e, Scherer2018e, Chen2020t}.

Beyond the valley picture derived from a square lattice geometry, our findings may also naturally carry over to hexagonal systems with two valleys at $\nu=K^{(\prime)}$. Examples include a variety of layered antiferromagnets~\cite{Feng2025f, Jiang2025s} as well as rhombohedral graphene multilayers (RMG). In RMG, the mean-field phase diagram~\cite{Koh2024s, Koh2024c} displays large regions of so-called spin valley locking (SVL, $\sigma^z\tau^z$), i.e., $f$-wave altermagnetism. Spin-polarized bands emerge despite the order paremeter's $f$-wave character since inversion symmetry is broken. Similar to the \sAM, it features a fully gapped Fermi surface. 
Very recently, a corresponding bilayer lattice model featuring collinear $f$-wave AM symmetry at high energies has been proposed~\cite{zhuang2025odd}. 

Finally, moiré heterostructures provide a highly tunable platform for realizing momentum-space non-symmorphic symmetries. While most studies have focused on $\Gamma$- or $K$-centered systems, recent work has identified a class of triangular-lattice moiré materials with low-energy states at the $M$ points \cite{Calugaru2025m}. The resulting continuum Hamiltonians exhibit emergent $k$-space non-symmorphic symmetries and realize projective representations of crystalline space groups, naturally matching the symmetry requirements underlying \sAMs. Importantly, an emergent momentum-space non-symmorphic mirror symmetry can render the low-energy states in a given valley effectively quasi-one-dimensional at the single-particle level, with strong dispersion along one direction and weak dispersion along the perpendicular one. In this sense, our 1D model~\cite{SM} can be viewed as a minimal toy realization of the same symmetry principle proposed for these $M$-point moiré systems. While preparing this manuscript, we became aware of recent works identifying explicit material platforms in twisted square-lattice moiré systems that realize analogous momentum-space translation symmetries, further broadening the scope of candidate systems \cite{bao2026moire, shi2026gvalley, eugenio2025tunable}.

\prlparagraph{Conclusion \& Outlook}%
Having proposed and analyzed the \sAM compensated magnetic state with spin-split bands, our work calls for a broader reconciliation of altermagnets from the viewpoint of their constituting symmetries. In order to eventually identify the most suited magnetic state for spintronics applications in terms of signal to noise ratio affected by quasiparticle poisoning and resilience to perturbations such as disorder, it appears worthwhile to expand the symmetry classification of magnetic states with spin-polarized magnon bands beyond just altermagnets based on crystal symmetry. Furthermore, the space of such magnetic states then needs to be investigated both from a localized moment~\cite{Smejkal2022, Smejkal2022a, zhao2025altermagnetism} and an itinerant~\cite{Durrnagel2025a, giuli2025altermagnetism, li2024dwave} perspective in order to reach a complete understanding of the classification and microscopic formation of unconventional quantum magnets. 

\bigskip

We thank A.~Fischer, J.~Schmalian, J.~Seufert, and B.~Trauzettel for fruitful discussions.
This research was funded by the Deutsche Forschungsgemeinschaft (DFG, German Research Foundation) – Project-ID 258499086 – SFB 1170; through the Würzburg-Dresden Cluster of Excellence on Complexity, Topology, and Dynamics in Quantum Materials (ctd.qmat) – Project-ID 390858490 – EXC 2147; and through the Research Unit QUAST – Project-ID 449872909 – FOR 5249. MD acknowledges support from the Studienstiftung des deutschen Volkes.

\let\oldaddcontentsline\addcontentsline
\renewcommand{\addcontentsline}[3]{}
\bibliography{references.bib,references_lieb.bib}
\let\addcontentsline\oldaddcontentsline

\supplement{Supplemental Material:\\
Extended {\itshape s}-wave altermagnets}

\tableofcontents

\section{Spin and valley polarization from mean-field and renormalization group calculations}
\label{SM:MF_RG}

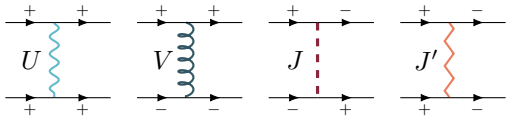
\begin{SCfigure}
  \begin{tikzpicture}[x=0.5cm, y=0.5cm, inner sep=0pt, outer sep=-0.5pt]
      \begin{scope}[shift={(7,0)}]
        \node[anchor=center] at (0.9,1) {$J$};
        \node (d1) at (.2,2) {};
        \node (d2) at (1.5,2) {};
        \node (d3) at (3.0,2) {};
        \node (c1) at (.2,0) {};
        \node (c2) at (1.5,0) {};
        \node (c3) at (3.0,0) {};
        \draw[pmmp] (d2) -- (c2);
        \draw[midarrow] (d1) -- node[vp,anchor=south]{} (d2);
        \draw[midarrow] (d2) -- node[vm,anchor=south]{} (d3);
        \draw[midarrow] (c1) -- node[vm,anchor=north]{} (c2);
        \draw[midarrow] (c2) -- node[vp,anchor=north]{} (c3);
      \end{scope}
    \begin{scope}[shift={(10.5,0)}]
        \node[anchor=center] at (0.9,1) {$J^\prime$};
        \node (d1) at (.2,2) {};
        \node (d2) at (1.5,2) {};
        \node (d3) at (3.0,2) {};
        \node (c1) at (.2,0) {};
        \node (c2) at (1.5,0) {};
        \node (c3) at (3.0,0) {};
        \draw[ppmm] (d2) -- (c2);
        \draw[midarrow] (d1) -- node[vp,anchor=south]{} (d2);
        \draw[midarrow] (d2) -- node[vm,anchor=south]{} (d3);
        \draw[midarrow] (c1) -- node[vp,anchor=north]{} (c2);
        \draw[midarrow] (c2) -- node[vm,anchor=north]{} (c3);
    \end{scope} 
    \begin{scope}[shift={(0, 0)}]
        \node[anchor=center] at (0.9,1) {$U$};
        \node (d1) at (.2,2) {};
        \node (d2) at (1.5,2) {};
        \node (d3) at (3.0,2) {};
        \node (c1) at (.2,0) {};
        \node (c2) at (1.5,0) {};
        \node (c3) at (3.0,0) {};
        \draw[pppp] (d2) -- (c2);
        \draw[midarrow] (d1) -- node[vp,anchor=south]{} (d2);
        \draw[midarrow] (d2) -- node[vp,anchor=south]{} (d3);
        \draw[midarrow] (c1) -- node[vp,anchor=north]{} (c2);
        \draw[midarrow] (c2) -- node[vp,anchor=north]{} (c3);
      \end{scope}
    \begin{scope}[shift={(3.5, 0)}]
        \node[anchor=center] at (0.9,1) {$V$};
        \node (d1) at (.2,2) {};
        \node (d2) at (1.5,2) {};
        \node (d3) at (3.0,2) {};
        \node (c1) at (.2,0) {};
        \node (c2) at (1.5,0) {};
        \node (c3) at (3.0,0) {};
        \draw[pmpm] (d2) -- (c2);
        \draw[midarrow] (d1) -- node[vp,anchor=south]{} (d2);
        \draw[midarrow] (d2) -- node[vp,anchor=south]{} (d3);
        \draw[midarrow] (c1) -- node[vm,anchor=north]{} (c2);
        \draw[midarrow] (c2) -- node[vm,anchor=north]{} (c3);
    \end{scope} 
\end{tikzpicture}
\hspace{2ex}
\caption{Diagrammatic representation of the bare interaction vertex according to \cref{eqn:bare_interaction}. The spin index is not explicitly written, but is conserved along all bare propagator lines.}
\label{fig:bare_interaction}
\end{SCfigure}

To analyze the propensity of an \sAM state in the given geometry, we consider a general two valley system with two equal spherical valleys $\tau = \pm$ around the $\Gamma$ and $M$ point as given by \cref{eqn:H0_valley} and we employ a two particle interaction
\begin{equation}
\begin{split}
    H_\text{I} =& \, \frac{1}{2} \sum_{\{\bvec k_i \}}
    ( U \tau^0_{\nu \nu^\prime} \sigma^x_{\sigma \sigma^\prime}
    + V \tau^x_{\nu \nu^\prime} ) \,
    \cre_{\bvec k_0, \nu, \sigma}
    \cre_{\bvec k_1, \nu^\prime, \sigma^\prime}
    \ann_{\bvec k_2, \nu^\prime, \sigma^\prime}
    \ann_{\bvec k_3, \nu \sigma}
    \\
    +& \, J \sum_{\{\bvec k_i \}} \tau^x_{\nu \nu^\prime} \, 
    \cre_{\bvec k_0, \nu, \sigma}
    \cre_{\bvec k_1, \nu^\prime, \sigma^\prime}
    \ann_{\bvec k_2, \nu, \sigma^\prime}
    \ann_{\bvec k_3, \nu^\prime, \sigma} \\
    +& \, J^\prime \sum_{\{\bvec k_i \}} \tau^x_{\nu \nu^\prime} \sigma^x_{\sigma \sigma^\prime} \, 
    \cre_{\bvec k_0, \nu, \sigma}
    \cre_{\bvec k_1, \nu, \sigma^\prime}
    \ann_{\bvec k_2, \nu^\prime, \sigma^\prime}
    \ann_{\bvec k_3, \nu^\prime, \sigma} 
\end{split}
\label{eqn:bare_interaction}
\end{equation}
consisting of the four symmetry allowed couplings: an intra-(inter-) valley repulsion $U$ ($V$) and an intervalley Hund's and pair hopping interaction $J$ and $J^\prime$, respectively~\cite{Chubukov2008m}.
A diagrammatic representation of \cref{eqn:bare_interaction} is given in \cref{fig:bare_interaction}.
Here we implicitly imply $\sum_{\{\bvec k_i \}} = \sum_{\nu \nu^\prime} \sum_{\sigma \sigma^\prime} \sum_{\bvec k_0 \bvec k_1 \bvec k_2 \bvec k_3} \delta(\bvec k_0 + \bvec k_1 - \bvec k_2 - \bvec k_3)$ to preserve momentum conservation and all repeated indices are summed over.

We analyze the system's ordering tendencies in close correspondence to Ref.~\cite{Raines2024t}.

\subsection{Symmetries and possible many-body ground states}
\label{SM:MF_RG_groundstates}

The possible spin and valley orders of the system are given by
\begin{itemize}
    \item valley polarization (VP): $\Delta \propto \tau^z \sigma^0$
    \item spin polarization (SP): $\Delta \propto \tau^0 \sigma^{x,y,z}$
    \item staggered spin polarisation (SSP): $\Delta \propto \tau^z \sigma^{x,y,z}$
    \item inter-valley coherence (IVC): $\Delta \propto \tau^{x,y} \sigma^0$
    \item spin inter valley coherence (SIVC): $\Delta \propto \tau^{x,y} \sigma^{x,y,z}$
\end{itemize}
The enforced symmetry between the two valleys dictates equal bare susceptibilities
\begin{equation}
    \chi^0_{\nu \nu^\prime}(\Gamma) = - \frac{1}{\beta} \sum_{\bvec k n} G^0_\nu(\bvec k + \bvec q, \omega_n) G^0_{\nu^\prime}(\bvec k, \omega_n) |_{\bvec q=\Gamma} \xrightarrow[\beta \rightarrow \infty] ~\rho \qquad \forall~\nu, \nu^\prime \,,
\label{eqn:chi_0}
\end{equation}
and the system features an $SU(4)$ symmetry.
Here $G^0_{\nu}(\bvec k, \omega_n)$ is the Green's function of a state in valley $\nu$, with momentum $\bvec k$ and fermionic Matsubara frquency $\omega_n = \frac{(2 n + 1) \pi}{\beta}$ and $\rho = \lim_{\beta \rightarrow \infty} \chi^0(\bvec q = 0)$ is the density of states at the Fermi level.

For $U = V$, $J = J^\prime = 0$, the system retains its $SU(4)$ symmetry and the mean-field (MF)/RPA susceptibilities of all states are equivalent
\begin{equation}
    \chi^\mathrm{MF} = \frac{4 \rho}{1 - \rho \Gamma}
\label{eqn:chi_MF}
\end{equation}
where $\Gamma = U$ for all aforementioned instabilities and the factor of $4$ stems from the $2 \times 2$ spin and valley degeneracy. 
In general, this $SU(4)$ symmetry is broken by interaction effects.

\subsection{Mean-field analysis}

Analyzing the bare interaction projected into the respective MF channels as performed in \cref{fig:MF_scattering}, we obtain different effective interactions for the various spin and valley orders given by
\begin{equation}
    \begin{split}
        \Gamma_\text{VP} =  \, 2V - U - J\\
        \Gamma_\text{SP} =  \, U + J\\
        \Gamma_\text{SSP} =  \, U - J\\
        \Gamma_\text{IVC} =  \, V - J - J^\prime \\
        \Gamma_\text{SIVC} =  \, V + J^\prime\\
    \end{split}
\label{eqn:MF_couplings}
\end{equation}
According to \cref{eqn:chi_MF}, the interacting susceptibility and therefore the critical temperature of the MF state is highest for the largest MF interaction value.
Hence, we see that the SSP state corresponding to the \sAM order is the favored FS instability for $U > V$ and $J < 0$.
We further note, that the SP and SSP state are only discriminated by the Hund's coupling $J$ and therefore degenerate for $J = 0$.

While the Hund's coupling usually has a positive sign, i.e., favors parallel spin alignment of different orbitals, it is worth noting that the couplings given in \cref{eqn:MF_couplings} are in valley space and represent an effective low energy model. To convert between the physical couplings in orbital space and the effective valley couplings, one has to express the Hubbard Kanamori vertex in new operators that diagonalize the kinetic Hamiltonian close to the Fermi level and expand their momentum structure up to the leading lattice harmonic~\cite{Chubukov2012pairing}. The effective couplings---including their signs---are consequently highly dependent on the relative phases of the Bloch states along the Fermi surface and the underlying microscopic Hamiltonian. Indeed, we will see in \cref{SM:FRG} that this in combination with renormalization effects from higher bands renders $J < 0$ for the model of \cref{latticemodel}.

\begin{SCfigure}
\begin{tikzpicture}[x=0.5cm, y=0.5cm, inner sep=0pt, outer sep=-0.5pt]
    \begin{scope}[shift={(0,0)}]
        \node[anchor=center] at (-.1,1) {$\uparrow$};
        \node[anchor=center] at (2.8,1) {$\uparrow$};
        \node (d1) at (.2,1) {};
        \node (d2) at (1.5,2) {};
        \node (d3) at (3.0,2) {};
        \node (c1) at (.2,1) {};
        \node (c2) at (1.5,0) {};
        \node (c3) at (3.0,0) {};
        \draw[pmpm] (d2) -- (c2);
        \pic [draw, fill=gray!60, angle eccentricity=1.2, angle radius=3mm,] {angle=c2--c1--d2};
        \draw[midarrow] (d1) -- node[vp,anchor=south]{} (d2);
        \draw[midarrow] (d2) -- node[vp,anchor=south]{} (d3);
        \draw[midarrow] (c2) -- node[vm,anchor=north]{} (c1);
        \draw[midarrow] (c3) -- node[vm,anchor=north]{} (c2);
    \end{scope}
    \begin{scope}[shift={(4.,0)}]
        \node[anchor=center] at (-.1,1) {$\uparrow$};
        \node[anchor=center] at (4.,1) {$\uparrow\downarrow$};
        \node (d1) at (.2,1) {};
        \node (d2) at (1.5,2.6) {};
        \node (d3) at (4.0,2) {};
        \node (c1) at (.2,1) {};
        \node (c2) at (1.5,-0.6) {};
        \node (c3) at (4.0,0) {};
        \node (m1) at (1.6,1) {};
        \node (m2) at (3.4,1) {};
        \draw[pmmp] (m1) -- (m2);
        \pic [draw, fill=gray!60, angle eccentricity=1.2, angle radius=3mm,] {angle=c2--c1--d2};
        \draw[midarrow] (d1) .. controls (d2) and (1.6, 2) .. node[vp,anchor=south]{} (m1);
        \draw[midarrow] (m1) .. controls (1.6, 0) and (c2).. node[vm,anchor=north]{}(c1);
        \draw[midarrow] (c3) .. controls (3.8,0) and (3.4, 0) .. node[vm,anchor=north]{} (m2);
        \draw[midarrow] (m2) .. controls (3.4, 2) and (3.8,2) .. node[vp,anchor=south]{} (d3);
    \end{scope}
    \begin{scope}[shift={(9.,0)}]
        \node[anchor=center] at (-.1,1) {$\uparrow$};
        \node[anchor=center] at (3.8,1) {$\downarrow$};
        \node (d1) at (.2,1) {};
        \node (d2) at (1.5,2.6) {};
        \node (d3) at (4.0,2) {};
        \node (c1) at (.2,1) {};
        \node (c2) at (1.5,-0.6) {};
        \node (c3) at (4.0,0) {};
        \node (m1) at (1.6,1) {};
        \node (m2) at (3.4,1) {};
        \draw[ppmm] (m1) -- (m2);
        \pic [draw, fill=gray!60, angle eccentricity=1.2, angle radius=3mm,] {angle=c2--c1--d2};
        \draw[midarrow] (d1) .. controls (d2) and (1.6, 2) .. node[vp,anchor=south]{} (m1);
        \draw[midarrow] (m1) .. controls (1.6, 0) and (c2).. node[vm,anchor=north]{}(c1);
        \draw[midarrow] (c3) .. controls (3.8,0) and (3.4, 0) .. node[vp,anchor=north]{} (m2);
        \draw[midarrow] (m2) .. controls (3.4, 2) and (3.8,2) .. node[vm,anchor=south]{} (d3);
    \end{scope}
    \begin{scope}[shift={(0,3.8)}]
    \begin{scope}[shift={(0,0)}]
        \node[anchor=center] at (-.1,1) {$\uparrow$};
        \node[anchor=center] at (2.8,1) {$\uparrow$};
        \node (d1) at (.2,1) {};
        \node (d2) at (1.5,2) {};
        \node (d3) at (3.0,2) {};
        \node (c1) at (.2,1) {};
        \node (c2) at (1.5,0) {};
        \node (c3) at (3.0,0) {};
        \draw[pmmp] (d2) -- (c2);
        \draw [line width=0.1mm] (c2) coordinate (A) -- (c1) coordinate (B) -- (d2) coordinate (C)
            pic [draw=blue, fill=gray!50, angle radius=3mm] {angle};
        \draw[midarrow] (d1) -- node[vp,anchor=south]{} (d2);
        \draw[midarrow] (d2) -- node[vm,anchor=south]{} (d3);
        \draw[midarrow] (c2) -- node[vp,anchor=north]{} (c1);
        \draw[midarrow] (c3) -- node[vm,anchor=north]{} (c2);
    \end{scope}
    \begin{scope}[shift={(4,0)}]
        \node[anchor=center] at (-.1,1) {$\uparrow$};
        \node[anchor=center] at (4.,1) {$\uparrow\downarrow$};
        \node (d1) at (.2,1) {};
        \node (d2) at (1.5,2.6) {};
        \node (d3) at (4.0,2) {};
        \node (c1) at (.2,1) {};
        \node (c2) at (1.5,-0.6) {};
        \node (c3) at (4.0,0) {};
        \node (m1) at (1.6,1) {};
        \node (m2) at (3.4,1) {};
        \draw[pmpm] (m1) -- (m2);
        \pic [draw, fill=gray!60, angle eccentricity=1.2, angle radius=3mm,] {angle=c2--c1--d2};
        \draw[midarrow] (d1) .. controls (d2) and (1.6, 2) .. node[vp,anchor=south]{} (m1);
        \draw[midarrow] (m1) .. controls (1.6, 0) and (c2).. node[vp,anchor=north]{}(c1);
        \draw[midarrow] (c3) .. controls (3.8,0) and (3.4, 0) .. node[vm,anchor=north]{} (m2);
        \draw[midarrow] (m2) .. controls (3.4, 2) and (3.8,2) .. node[vm,anchor=south]{} (d3);
    \end{scope}
    \begin{scope}[shift={(9.,0)}]
        \node[anchor=center] at (-.1,1) {$\uparrow$};
        \node[anchor=center] at (3.8,1) {$\downarrow$};
        \node (d1) at (.2,1) {};
        \node (d2) at (1.5,2.6) {};
        \node (d3) at (4.0,2) {};
        \node (c1) at (.2,1) {};
        \node (c2) at (1.5,-0.6) {};
        \node (c3) at (4.0,0) {};
        \node (m1) at (1.6,1) {};
        \node (m2) at (3.4,1) {};
        \draw[pmmp] (m1) -- (m2);
        \pic [draw, fill=gray!60, angle eccentricity=1.2, angle radius=3mm,] {angle=c2--c1--d2};
        \draw[midarrow] (d1) .. controls (d2) and (1.6, 2) .. node[vp,anchor=south]{} (m1);
        \draw[midarrow] (m1) .. controls (1.6, 0) and (c2).. node[vp,anchor=north]{}(c1);
        \draw[midarrow] (c3) .. controls (3.8,0) and (3.4, 0) .. node[vp,anchor=north]{} (m2);
        \draw[midarrow] (m2) .. controls (3.4, 2) and (3.8,2) .. node[vp,anchor=south]{} (d3);
    \end{scope}
    \end{scope}
\end{tikzpicture}
\hspace{2ex}
\caption{Scattering elements for different particle hole pairs with well-defined spin. The line shapes indicate the bare vertex elements depicted in \cref{fig:bare_interaction}.}
\label{fig:ph_scattering}
\end{SCfigure}
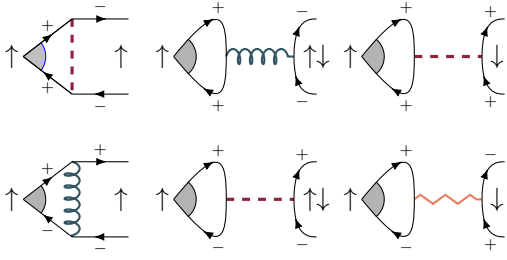

\begin{SCfigure}
\begin{tikzpicture}[x=0.5cm, y=0.5cm, inner sep=0pt, outer sep=-0.5pt]
    \begin{scope}[shift={(0,4 * 3.8)}]
    \begin{scope}[shift={(0,0)}]
        \node[anchor=center] at (-.5,1) {$\Delta_{\text{VP}}$};
        \node[anchor=center] at (2.3,1) {$\Gamma_{\text{VP}}$};
        \node (d1) at (.2,1) {};
        \node (d2) at (1.5,2) {};
        \node (d3) at (3.0,2) {};
        \node (c1) at (.2,1) {};
        \node (c2) at (1.5,0) {};
        \node (c3) at (3.0,0) {};
        \draw[mf] (d2) -- (c2);
        \pic [draw, fill=gray!60, angle eccentricity=1.2, angle radius=3mm,] {angle=c2--c1--d2};
        \draw[midarrow] (d1) -- (d2);
        \draw[midarrow] (d2) -- (d3);
        \draw[midarrow] (c2) -- (c1);
        \draw[midarrow] (c3) -- (c2);
    \end{scope}
    \begin{scope}[shift={(4.5,0)}]
        \node[anchor=center] at (-.5,1) {$= -$};
        \node (d1) at (.2,1) {};
        \node (d2) at (1.5,2) {};
        \node (d3) at (3.0,2) {};
        \node (c1) at (.2,1) {};
        \node (c2) at (1.5,0) {};
        \node (c3) at (3.0,0) {};
        \draw[pmmp] (d2) -- (c2);
        \pic [draw, fill=gray!60, angle eccentricity=1.2, angle radius=3mm,] {angle=c2--c1--d2};
        \draw[midarrow] (d1) -- (d2);
        \draw[midarrow] (d2) -- (d3);
        \draw[midarrow] (c2) -- (c1);
        \draw[midarrow] (c3) -- (c2);
    \end{scope}
    \begin{scope}[shift={(8.5,0)}]
        \node[anchor=center] at (-.5,1) {$- 2 $};
        \node (d1) at (.2,1) {};
        \node (d2) at (1.5,2.6) {};
        \node (d3) at (4.0,2) {};
        \node (c1) at (.2,1) {};
        \node (c2) at (1.5,-0.6) {};
        \node (c3) at (4.0,0) {};
        \node (m1) at (1.6,1) {};
        \node (m2) at (3.4,1) {};
        \draw[pmpm] (m1) -- (m2);
        \pic [draw, fill=gray!60, angle eccentricity=1.2, angle radius=3mm,] {angle=c2--c1--d2};
        \draw[midarrow] (d1) .. controls (d2) and (1.6, 2) .. (m1);
        \draw[midarrow] (m1) .. controls (1.6, 0) and (c2).. (c1);
        \draw[midarrow] (c3) .. controls (3.8,0) and (3.4, 0) .. (m2);
        \draw[midarrow] (m2) .. controls (3.4, 2) and (3.8,2) .. (d3);
    \end{scope}
    \begin{scope}[shift={(13.5,0)}]
        \node[anchor=center] at (-.5,1) {$+$};
        \node (d1) at (.2,1) {};
        \node (d2) at (1.5,2.6) {};
        \node (d3) at (4.0,2) {};
        \node (c1) at (.2,1) {};
        \node (c2) at (1.5,-0.6) {};
        \node (c3) at (4.0,0) {};
        \node (m1) at (1.6,1) {};
        \node (m2) at (3.4,1) {};
        \draw[pppp] (m1) -- (m2);
        \pic [draw, fill=gray!60, angle eccentricity=1.2, angle radius=3mm,] {angle=c2--c1--d2};
        \draw[midarrow] (d1) .. controls (d2) and (1.6, 2) .. (m1);
        \draw[midarrow] (m1) .. controls (1.6, 0) and (c2).. (c1);
        \draw[midarrow] (c3) .. controls (3.8,0) and (3.4, 0) .. (m2);
        \draw[midarrow] (m2) .. controls (3.4, 2) and (3.8,2) .. (d3);
    \end{scope}
    \end{scope}
    
    \begin{scope}[shift={(0,3 * 3.8)}]
    \begin{scope}[shift={(0,0)}]
        \node[anchor=center] at (-.5,1) {$\Delta_{\text{SP}}$};
        \node[anchor=center] at (2.3,1) {$\Gamma_{\text{SP}}$};
        \node (d1) at (.2,1) {};
        \node (d2) at (1.5,2) {};
        \node (d3) at (3.0,2) {};
        \node (c1) at (.2,1) {};
        \node (c2) at (1.5,0) {};
        \node (c3) at (3.0,0) {};
        \draw[mf] (d2) -- (c2);
        \pic [draw, fill=gray!60, angle eccentricity=1.2, angle radius=3mm,] {angle=c2--c1--d2};
        \draw[midarrow] (d1) -- (d2);
        \draw[midarrow] (d2) -- (d3);
        \draw[midarrow] (c2) -- (c1);
        \draw[midarrow] (c3) -- (c2);
    \end{scope}
    \begin{scope}[shift={(4.5,0)}]
        \node[anchor=center] at (-.9,1) {$=$};
        \node (d1) at (.2,1) {};
        \node (d2) at (1.5,2) {};
        \node (d3) at (3.0,2) {};
        \node (c1) at (.2,1) {};
        \node (c2) at (1.5,0) {};
        \node (c3) at (3.0,0) {};
        \draw[pmmp] (d2) -- (c2);
        \pic [draw, fill=gray!60, angle eccentricity=1.2, angle radius=3mm,] {angle=c2--c1--d2};
        \draw[midarrow] (d1) -- (d2);
        \draw[midarrow] (d2) -- (d3);
        \draw[midarrow] (c2) -- (c1);
        \draw[midarrow] (c3) -- (c2);
    \end{scope}
    \begin{scope}[shift={(8.5,0)}]
        \node[anchor=center] at (-.5,1) {$-$};
        \node (d1) at (.2,1) {};
        \node (d2) at (1.5,2.6) {};
        \node (d3) at (4.0,2) {};
        \node (c1) at (.2,1) {};
        \node (c2) at (1.5,-0.6) {};
        \node (c3) at (4.0,0) {};
        \node (m1) at (1.6,1) {};
        \node (m2) at (3.4,1) {};
        \draw[pppp] (m1) -- (m2);
        \pic [draw, fill=gray!60, angle eccentricity=1.2, angle radius=3mm,] {angle=c2--c1--d2};
        \draw[midarrow] (d1) .. controls (d2) and (1.6, 2) .. (m1);
        \draw[midarrow] (m1) .. controls (1.6, 0) and (c2).. (c1);
        \draw[midarrow] (c3) .. controls (3.8,0) and (3.4, 0) .. (m2);
        \draw[midarrow] (m2) .. controls (3.4, 2) and (3.8,2) .. (d3);
    \end{scope}
    \end{scope}
    
    \begin{scope}[shift={(0,2 * 3.8)}]
    \begin{scope}[shift={(0,0)}]
        \node[anchor=center] at (-.7,1) {$\Delta_{\text{SSP}}$};
        \node[anchor=center] at (2.45,1) {$\Gamma_{\text{SSP}}$};
        \node (d1) at (.2,1) {};
        \node (d2) at (1.5,2) {};
        \node (d3) at (3.0,2) {};
        \node (c1) at (.2,1) {};
        \node (c2) at (1.5,0) {};
        \node (c3) at (3.0,0) {};
        \draw[mf] (d2) -- (c2);
        \pic [draw, fill=gray!60, angle eccentricity=1.2, angle radius=3mm,] {angle=c2--c1--d2};
        \draw[midarrow] (d1) -- (d2);
        \draw[midarrow] (d2) -- (d3);
        \draw[midarrow] (c2) -- (c1);
        \draw[midarrow] (c3) -- (c2);
    \end{scope}
    \begin{scope}[shift={(4.5,0)}]
        \node[anchor=center] at (-.5,1) {$= -$};
        \node (d1) at (.2,1) {};
        \node (d2) at (1.5,2) {};
        \node (d3) at (3.0,2) {};
        \node (c1) at (.2,1) {};
        \node (c2) at (1.5,0) {};
        \node (c3) at (3.0,0) {};
        \draw[pmmp] (d2) -- (c2);
        \pic [draw, fill=gray!60, angle eccentricity=1.2, angle radius=3mm,] {angle=c2--c1--d2};
        \draw[midarrow] (d1) -- (d2);
        \draw[midarrow] (d2) -- (d3);
        \draw[midarrow] (c2) -- (c1);
        \draw[midarrow] (c3) -- (c2);
    \end{scope}
    \begin{scope}[shift={(8.5,0)}]
        \node[anchor=center] at (-.5,1) {$-$};
        \node (d1) at (.2,1) {};
        \node (d2) at (1.5,2.6) {};
        \node (d3) at (4.0,2) {};
        \node (c1) at (.2,1) {};
        \node (c2) at (1.5,-0.6) {};
        \node (c3) at (4.0,0) {};
        \node (m1) at (1.6,1) {};
        \node (m2) at (3.4,1) {};
        \draw[pppp] (m1) -- (m2);
        \pic [draw, fill=gray!60, angle eccentricity=1.2, angle radius=3mm,] {angle=c2--c1--d2};
        \draw[midarrow] (d1) .. controls (d2) and (1.6, 2) .. (m1);
        \draw[midarrow] (m1) .. controls (1.6, 0) and (c2).. (c1);
        \draw[midarrow] (c3) .. controls (3.8,0) and (3.4, 0) .. (m2);
        \draw[midarrow] (m2) .. controls (3.4, 2) and (3.8,2) .. (d3);
    \end{scope}
    \end{scope}

    \begin{scope}[shift={(0,1 * 3.8)}]
    \begin{scope}[shift={(0,0)}]
        \node[anchor=center] at (-.6,1) {$\Delta_{\text{IVC}}$};
        \node[anchor=center] at (2.4,1) {$\Gamma_{\text{IVC}}$};
        \node (d1) at (.2,1) {};
        \node (d2) at (1.5,2) {};
        \node (d3) at (3.0,2) {};
        \node (c1) at (.2,1) {};
        \node (c2) at (1.5,0) {};
        \node (c3) at (3.0,0) {};
        \draw[mf] (d2) -- (c2);
        \pic [draw, fill=gray!60, angle eccentricity=1.2, angle radius=3mm,] {angle=c2--c1--d2};
        \draw[midarrow] (d1) -- (d2);
        \draw[midarrow] (d2) -- (d3);
        \draw[midarrow] (c2) -- (c1);
        \draw[midarrow] (c3) -- (c2);
    \end{scope}
    \begin{scope}[shift={(4.5,0)}]
        \node[anchor=center] at (-.9,1) {$=$};
        \node (d1) at (.2,1) {};
        \node (d2) at (1.5,2) {};
        \node (d3) at (3.0,2) {};
        \node (c1) at (.2,1) {};
        \node (c2) at (1.5,0) {};
        \node (c3) at (3.0,0) {};
        \draw[pmpm] (d2) -- (c2);
        \pic [draw, fill=gray!60, angle eccentricity=1.2, angle radius=3mm,] {angle=c2--c1--d2};
        \draw[midarrow] (d1) -- (d2);
        \draw[midarrow] (d2) -- (d3);
        \draw[midarrow] (c2) -- (c1);
        \draw[midarrow] (c3) -- (c2);
    \end{scope}
    \begin{scope}[shift={(8.5,0)}]
        \node[anchor=center] at (-.5,1) {$+$};
        \node (d1) at (.2,1) {};
        \node (d2) at (1.5,2.6) {};
        \node (d3) at (4.0,2) {};
        \node (c1) at (.2,1) {};
        \node (c2) at (1.5,-0.6) {};
        \node (c3) at (4.0,0) {};
        \node (m1) at (1.6,1) {};
        \node (m2) at (3.4,1) {};
        \draw[pmmp] (m1) -- (m2);
        \pic [draw, fill=gray!60, angle eccentricity=1.2, angle radius=3mm,] {angle=c2--c1--d2};
        \draw[midarrow] (d1) .. controls (d2) and (1.6, 2) .. (m1);
        \draw[midarrow] (m1) .. controls (1.6, 0) and (c2).. (c1);
        \draw[midarrow] (c3) .. controls (3.8,0) and (3.4, 0) .. (m2);
        \draw[midarrow] (m2) .. controls (3.4, 2) and (3.8,2) .. (d3);
    \end{scope}
    \begin{scope}[shift={(13.5,0)}]
        \node[anchor=center] at (-.5,1) {$+$};
        \node (d1) at (.2,1) {};
        \node (d2) at (1.5,2.6) {};
        \node (d3) at (4.0,2) {};
        \node (c1) at (.2,1) {};
        \node (c2) at (1.5,-0.6) {};
        \node (c3) at (4.0,0) {};
        \node (m1) at (1.6,1) {};
        \node (m2) at (3.4,1) {};
        \draw[ppmm] (m1) -- (m2);
        \pic [draw, fill=gray!60, angle eccentricity=1.2, angle radius=3mm,] {angle=c2--c1--d2};
        \draw[midarrow] (d1) .. controls (d2) and (1.6, 2) .. (m1);
        \draw[midarrow] (m1) .. controls (1.6, 0) and (c2).. (c1);
        \draw[midarrow] (c3) .. controls (3.8,0) and (3.4, 0) .. (m2);
        \draw[midarrow] (m2) .. controls (3.4, 2) and (3.8,2) .. (d3);
    \end{scope}
    \end{scope}

    \begin{scope}[shift={(0,0 * 3.8)}]
    \begin{scope}[shift={(0,0)}]
        \node[anchor=center] at (-.7,1) {$\Delta_{\text{SIVC}}$};
        \node[anchor=center] at (2.5,1) {$\Gamma_{\text{SIVC}}$};
        \node (d1) at (.2,1) {};
        \node (d2) at (1.5,2) {};
        \node (d3) at (3.0,2) {};
        \node (c1) at (.2,1) {};
        \node (c2) at (1.5,0) {};
        \node (c3) at (3.0,0) {};
        \draw[mf] (d2) -- (c2);
        \pic [draw, fill=gray!60, angle eccentricity=1.2, angle radius=3mm,] {angle=c2--c1--d2};
        \draw[midarrow] (d1) -- (d2);
        \draw[midarrow] (d2) -- (d3);
        \draw[midarrow] (c2) -- (c1);
        \draw[midarrow] (c3) -- (c2);
    \end{scope}
    \begin{scope}[shift={(4.5,0)}]
        \node[anchor=center] at (-.5,1) {$=$};
        \node (d1) at (.2,1) {};
        \node (d2) at (1.5,2) {};
        \node (d3) at (3.0,2) {};
        \node (c1) at (.2,1) {};
        \node (c2) at (1.5,0) {};
        \node (c3) at (3.0,0) {};
        \draw[pmpm] (d2) -- (c2);
        \pic [draw, fill=gray!60, angle eccentricity=1.2, angle radius=3mm,] {angle=c2--c1--d2};
        \draw[midarrow] (d1) -- (d2);
        \draw[midarrow] (d2) -- (d3);
        \draw[midarrow] (c2) -- (c1);
        \draw[midarrow] (c3) -- (c2);
    \end{scope}
    \begin{scope}[shift={(8.5,0)}]
        \node[anchor=center] at (-.5,1) {$-$};
        \node (d1) at (.2,1) {};
        \node (d2) at (1.5,2.6) {};
        \node (d3) at (4.0,2) {};
        \node (c1) at (.2,1) {};
        \node (c2) at (1.5,-0.6) {};
        \node (c3) at (4.0,0) {};
        \node (m1) at (1.6,1) {};
        \node (m2) at (3.4,1) {};
        \draw[ppmm] (m1) -- (m2);
        \pic [draw, fill=gray!60, angle eccentricity=1.2, angle radius=3mm,] {angle=c2--c1--d2};
        \draw[midarrow] (d1) .. controls (d2) and (1.6, 2) .. (m1);
        \draw[midarrow] (m1) .. controls (1.6, 0) and (c2).. (c1);
        \draw[midarrow] (c3) .. controls (3.8,0) and (3.4, 0) .. (m2);
        \draw[midarrow] (m2) .. controls (3.4, 2) and (3.8,2) .. (d3);
    \end{scope}
    \end{scope}
\end{tikzpicture}
\caption{Scattering elements contributing to the different MF channels according to the ph scattering elements depicted in \cref{fig:ph_scattering}. All bare propagator lines have a well-defined spin and we have indicated spin degeneracies explicitly by factors of $2$. The spin and valley index of the different individual propagators are fixed by the definition of the MF order parameter and the interaction vertex in \cref{fig:bare_interaction}.}
\label{fig:MF_scattering}
\end{SCfigure}
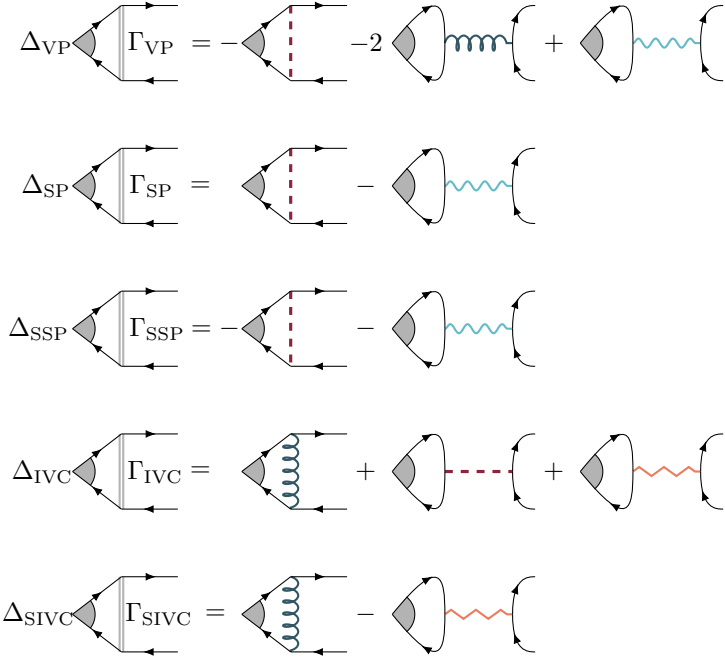

\subsection{One-loop RG analysis}

\begin{figure}
    \centering
    \includegraphics[width=\textwidth]{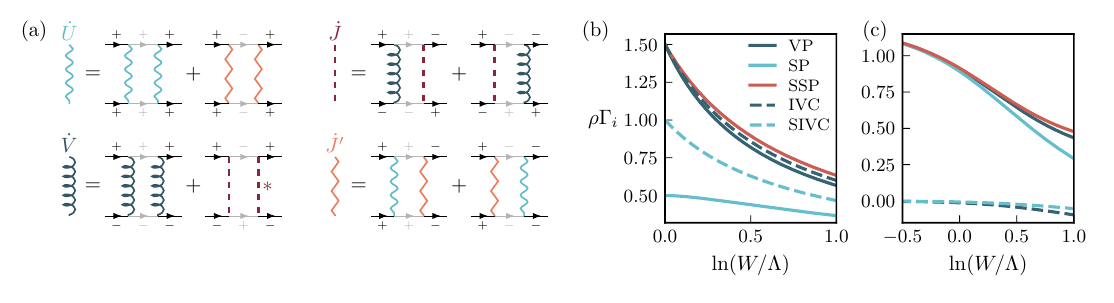}    
\caption{Renormalization of the interaction parameter to logarithmic accuracy.
(a)~one-loop particle particle renormalization of the initial couplings given in \cref{fig:bare_interaction}. The light gray lines represent derivatives of the pp bubble with respect to the cutoff $\Lambda$.
(b)~Exemplary flow of the MF couplings as a function of the low energy cutoff $\Lambda$. We have chosen $\rho U = \rho V = 1$, $J^\prime = 0$ and $J/U = -0.5$.
(c)~Flow of the MF couplings for the lattice model defined in \cref{latticemodel} using unbiased FRG calculations at $\rho U = 1.2$ and $V = J = J^\prime = 0$.}
\label{fig:flow}
\end{figure}

We can extend the analysis of interaction driven phenomena in the two valley system by considering renormalization effects of the interaction parameters beyond the RPA/MF level.
In general the renormalization of the couplings are given by the 1-loop RG eqauutions presented in Ref.~\cite{Raines2024t}.
Since both pockets are of electron type, the bare particle-hole (ph) susceptibility in \cref{eqn:chi_0} remains finite down to the lowest energy scales, while the corresponding particle-particle (pp) bubble features the generic Cooper log instability
\begin{equation}
    \chi^\text{pp}_{\nu \nu^\prime}(\bvec q = \Gamma) = \frac{1}{\beta} \sum_{\bvec k n} G^0_\nu(\bvec k, \omega_n) G^0_{\nu^\prime}(-\bvec k, -\omega_n) \rightarrow -\rho \ln \left(\frac{W}{\Lambda} \right)
\label{eqn:cooper_log}
\end{equation}
for $T = 0$, where $W$ is the bandwidth of the system and $\Lambda$ is an arbitrary long-distance (i.e. infrared) cutoff.
Hence, to logarithmic accuracy, we can focus on the diagrams, that feature a pp bubble in the perturbative vertex expansion.
We can now use $\Lambda$ as a control parameter to determine the renormalization of the initial couplings in \cref{eqn:bare_interaction} under a change of $\Lambda$ as we successively lower the scale from $W$ to the Fermi level.
To logarithmic accuracy the 1-loop RG equations are given by (compare \cref{fig:flow} and the notation in Ref.~\cite{Chubukov2008m})
\begin{equation}
\begin{split}
    \dot{U} = & \, - \, (U^2 + {J^\prime}^2) \,, \\
    \dot{V} = & \, - \, (V^2 + J^2) \,, \\
    \dot{J} = & \, - \, 2 V J \,, \\
    \dot{J^\prime} = & \, -\, 2 U J^\prime  \,,
\end{split}
\label{eqn:RG}
\end{equation}
with $\dot{\Gamma} = \frac{\text{d} \Gamma}{\text{d} \nu \ln(W/\Lambda)}$.

We note that for $J = J^\prime = 0$ the renormalization of the remaining couplings decouples and can be independently resummed in a geometric series as done in Ref.~\cite{Raines2024t, Calvera2025t}.
For finite $J, J^\prime$ this is not possible.

In general we can split the flow of the couplings in two regimes~\cite{Maiti2010r}:
For $\Lambda > E_\mathrm{F}$ all couplings and thereby interactions in the different MF channels couple with each other governed by \cref{eqn:RG}.
Below $E_\mathrm{F}$ the different MF channels decouple, since the valley structure of the external momenta limits the possible scattering channels. Thereby, the RG flow reduces to a ladder series in the respective channels that can be resummed down to zero energy resulting in the MF susceptibility of \cref{eqn:chi_MF} with $\Gamma_i = \Gamma_i^{E_\mathrm{F}}$ as obtained from the RG flow down to $E_\mathrm{F}$.
From the Stoner criterion for a phase transition $\rho \Gamma_i^{E_\mathrm{F}} = 1$, we can infer the physically realized instability as the one with the largest effective interaction at $E_\mathrm{F}$.
It is crucial to notice, that all investigated orders are not generic weak-coupling instabilities of the two valley system since the absence of ph nesting requires a finite bare interaction strength for them to emerge.

Since the renormalization is only due to particle-particle diagrams, it is possible to analytically integrate the RG flow down to $E_\mathrm{F}$.
Expressing the bare interaction in spinor basis for particle-particle bilinears $\vec{\Psi}^{\dag}_{\bvec k, \bvec k^\prime} = \big( \cre_{\bvec k,+} \cre_{\bvec k^\prime,+}, \cre_{\bvec k,-} \cre_{\bvec k^\prime,-}, \cre_{\bvec k,+} \cre_{\bvec k^\prime,-}, \cre_{\bvec k, -} \cre_{\bvec k^\prime, +} \big)$, $H_\text{I}$ acquires a block diagonal structure in valley space
\begin{equation}
    H_\text{I} = \sum_{\{\bvec k_i\}} \vec \Psi^{\dag}_{\bvec k_0, \bvec k_1}
    \begin{pmatrix}
        U & J^\prime & 0 & 0 \\
        J^\prime & U & 0 & 0 \\
        0 & 0 & V & J \\
        0 & 0 & J & V 
    \end{pmatrix}
    \vec\Psi^{\vphantom{\dag}}_{\bvec k_2, \bvec k_3}
     = \sum_{\{\bvec k_i\}} \vec \Psi^{\dag}_{\bvec k_0, \bvec k_1} \Gamma_0
    \vec \Psi^{\vphantom{\dag}}_{\bvec k_2, \bvec k_3} \,,
\end{equation}
and the Bethe-Salpeter equation gives
\begin{equation}
    \tilde \Gamma_0 =  \Gamma_0 + \Gamma_0 \chi^{pp}(\Gamma) \tilde \Gamma_0 \Rightarrow
    \tilde \Gamma_0 = (1 - \Gamma_0 \chi^{pp}(\Gamma) )^{-1} \Gamma_0 \ .
\end{equation}
Inserting \cref{eqn:cooper_log} we obtain at $\Lambda$ the renormalized couplings
\begin{equation}
\begin{split}
    U^\Lambda = & \, \frac{ U + \rho \ln(W / \Lambda) (U^2 - {J^\prime}^2)}{\big[ 1 + \rho \ln(W/\Lambda) U \big]^2 - \rho^2 \ln(W/\Lambda)^2 {J^\prime}^2} \,, \\
    V^\Lambda = & \, \frac{ V + \rho \ln(W / \Lambda) (V^2 - J^2)}{\big[ 1 + \rho \ln(W/\Lambda) V \big]^2 - \rho^2 \ln(W/\Lambda)^2 J^2} \,, \\
    J^\Lambda = & \, \frac{ J}{\big[ 1 + \rho \ln(W/\Lambda) V \big]^2 - \rho^2 \ln(W/\Lambda)^2 J^2} \,, \\
    {J'}^\Lambda = & \frac{J^\prime}{\big[ 1 + \rho \ln(W/\Lambda) U \big]^2 - \rho^2 \ln(W/\Lambda)^2 {J^\prime}^2} \,.
\end{split}
\label{eqn:renormalized_interactions}
\end{equation}
For $J = J^\prime = 0$ we recover the result presented in Ref.~\cite{Raines2024t, Calvera2025t}.
We can directly see, that for the generic setting $U, V > J,J^\prime$ the interaction strength is reduced by screening processes in the pp channel.
Since there is no native divergence in the ph channel, this renormalization is weak for sizable $\Lambda/W$ and does not effect the hierarchy of orders compared to the MF level except when they represent almost degenerate MF solutions.

We hence concentrate on bare interaction settings, that feature a degenerate MF solution involving the SSP state and inspect the splitting of the couplings induced by our 1-loop RG analysis:
\begin{itemize}
    \item  For $U = V > J^\prime = 0$, the system still features a degeneracy between SSP, VP and IVC on the MF level according to \cref{eqn:MF_couplings}. This degeneracy is broken when replacing the bare interaction parameters in \cref{eqn:MF_couplings} with the renormalized values from \cref{eqn:renormalized_interactions} and the SSP state is indeed the only leading instability beyond MF (cf. \cref{fig:flow}.
    \item Analogously, SP and SSP are degenerate for $U > V, J = 0$ and remain degenerate through the full RG flow since no $J$ can be  dynamically degenerated during the flow in leading logarithmic order. The same holds true for the $J = J^\prime = 0$ case.
\end{itemize}
Hence, the RG is only relevant in the $U = V$ case, where it singles out the SSP over VP and IVC.

\subsection{Functional renormalization group analysis}
\label{SM:FRG}

Turning from the effective valley model to the lattice model defined by \cref{latticemodel} and adding electronic correlations via an onsite interaction term $H_I = U \sum_{il} c^{\dagger}_{il\uparrow}c^{\vdagger}_{il\uparrow} \, c^{\dagger}_{il\downarrow} c^{\vdagger}_{il\downarrow}$ [$l$ denotes layer, corresponding to the vector degree of freedom in \cref{latticemodel}], we analyze the renormalization of the mean field coupling strength using unbiased functional renormalization group (FRG) calculations as implemented in the truncated unity backend of the divERGe library~\cite{Profe2024a, Profe2024b}.
Unlike the preceding 1-loop RG analysis the FRG takes not only processes up to leading logarithmic order into account, but incorporates all 1-particle irreducible diagrams of the three diagrammatic channels in the effective vertex renormalization.

To be comparable with the previous 1-loop RG analysis, we perform FRG calculations taking into account only intra-unit cell formfactors and evaluate the fermionic loop integrals on a $1000 \times 1000$ grid in the Brillouin zone. To include also the renormalization effects of the higher anergy bands, we take $\Lambda = 50 t_\parallel$ as initial value for the sharp frequency cutoff.
We analyze the MF couplings by calculating the expectation value of the MF groundstates $\Delta$ given in \cref{SM:MF_RG_groundstates} with the effective two particle vertex of the FRG $\Gamma^\Lambda$ at each flow step $\Gamma_i = \bra{\Delta_i} \Gamma^\Lambda \ket{\Delta_i}$.  

We exemplarily show the flow of the MF couplings in \cref{fig:flow}(c) for the kinetic parameter of \cref{fig:bandplot} at vanishing magnetization and take $U = 80 t_\parallel$ which corresponds to $\rho U = 1.2$. 
In accordance with the previous 1-loop RG analysis, all MF couplings decrease during the flow due to the renormalizing effect of the pairing channel. Still the SSP remains the leading instability throughout the flow and eventually diverges by means of a Stoner instability \cref{eqn:chi_MF} at sufficiently large bare interaction.
Notably, this suggests an effective valley Hund's coupling $J < 0$. This can be traced back to the structure of the microscopic Hamiltonian: Due to the strong orbital polarization of the valleys, they can be associated with one of the two sublattice sites in the unit cell. Then $J < 0$ corresponds to an antiferromagnetic nearest neighbor coupling that is mediated by a superexchange mechanism on the order $t_\perp^2 / U$.

In a full FRG treatment at lowered temperatures, other instabilities like antiferromagnetic order or valley singlet superconducting orders become competitive. We delegate a comprehensive study of the many-body phase diagram of \cref{latticemodel} to future work.

\section{Spin transport across an \sAM-N junction}
\label{SM:transport}

In this section we calculate the spin-dependent transport in a heterostructure setup depicted in \cref{fig:junction}:
One side of the junction is composed of the two-valley system discussed in the main part of the paper. It features two circular pockets around the BZ center $\Gamma$ and corner $M$ that are related by a chiral symmetry. Additionally we impose a spin-valley polarization that we incorporate on the mean-field level such that, i.e., $E_\uparrow(\bvec k) = E_\downarrow(\bvec k + M)$.
The low energy effective Hamiltonian is presented in \cref{eqn:H0_valley} in the main text.
The other side of the junction features a single spin-degenerate parabolic dispersion around $\Gamma$ as a stereotypical semi-metal bandstructure.
Since the pockets are well separated in momentum space and cannot mix, we can calculate their contribution to electronic transport separately.

We will start in the following with the pocket around $\Gamma$ and use the technique presented in Ref.~\cite{Breunig2021d} to obtain the current across the \sAM-N junction:
To model the low energy effective theory of the valley system presented in the main text, we consider a normal state Hamiltonian
\begin{equation}
    H_0 = \frac{k_x^2 + k_y^2}{2 m} \,,
\end{equation}
whose eigenstates with energy $E$ are given by the solutions of the static Schrödinger equation $H_\text{N} \psi = E \psi$. Here, as in the full paper we have set $\hbar = 1$.
We consider a junction with an interface between the altermagnetic (\sAM) and normal conducting (N) domain in the $x=0$ plane on a torus. Hence, spin $s = \pm 1$ for $\uparrow/\downarrow$ and the momentum in $y$-direction remain as good quantum numbers and $m$ and $\Ef$ are discontinuous across the interface:
\begin{equation}
    \begin{split}
        \Ef^\sigma(x) = & \, (\Ef^\text{\sAM} + \sigma \Delta ) \, \Theta(-x) + \Ef^\text{N} \, \Theta(x) \,, \\
        m(x) = & \, m_\text{\sAM} \, \Theta(-x) + m_\text{N} \, \Theta(x) \,.
    \end{split}
\end{equation}
$\Delta$ gives the size of the spin valley polarization.
Far away from the interface, we can neglect the effect of the boundary and calculate the electronic eigenspectrum
\begin{equation}
    E_{\tiny\substack{x \ll 0\\x \gg 0}} = \frac{k_x^2 + k_y^2}{2 m_\text{\sAM/N}} - \Ef^\text{\sAM/N}\,,
\end{equation}
alongside the corresponding eigenstates for left and right moving particles in mixed representation $\psi^{s\pm}(x, k_y) = \e^{\I k^\sigma(x) x}$ with
\begin{equation}
\begin{split}
    k^\sigma(x) = & \, \kappa^\sigma(x) \sqrt{1 \pm \frac{E}{\Ef^\sigma(x)} - \left( \frac{k_y}{\kappa^\sigma(x)} \right)} \,, \\
    \kappa^\sigma(x) = & \, \sqrt{2 m(x) \Ef^\sigma(x)} \,.
\end{split}
\end{equation}
The transmission and backscattering of a left incident plane wave solution at the junction is described by the total wavefunction
\begin{equation}
    \phi_\sigma(x) = \begin{cases}
        \psi_\text{\sAM}^{s+}(x) + a_\sigma \, \psi_\text{\sAM}^{s-}(x) \\
        b_\sigma \, \psi_\text{N}^{s+}(x)
    \end{cases} \ ,
\end{equation}
that fulfills the connectivity conditions at $x = 0$
\begin{equation}
    \begin{split}
        \lim_{\epsilon \rightarrow 0} \, ( \phi_\sigma(0 + \epsilon) - \phi_\sigma(0 - \epsilon) ) = {}& 0 \,, \\
        \lim_{\epsilon \rightarrow 0} \, \left( \frac{\phi^\prime_\sigma(0 + \epsilon)}{m_\text{N}} - \frac{\phi^\prime_\sigma(0 - \epsilon)}{m_\text{\sAM}} \right) = {}& 2 Z \phi_\sigma(0) \,.
    \end{split}
\end{equation}
Here, $Z$ acts as an elastic scattering potential at the interface and spin is conserved across the junction. Thereby, the energy eigenvalue $E$ is a quantum number for the global wavefunction $\phi(x)$, too.
Inserting the plane wave solutions from above one obtains the spin dependent transmission coefficient for the propagating modes with group velocities $v_i = k_i / m_i$~\cite{Breunig2021d}:
\begin{equation}
    T_\sigma(E, k_y) = 1 - |a_\sigma|^2 = \frac{
    4 v^\sigma_\text{\sAM} v_\text{N} \Theta(\kappa^\sigma_\text{\sAM} - |k_y|) \Theta(\kappa_\text{N} - |k_y|)}
    {4 Z^2 + (v^\sigma_\text{\sAM} + v_\text{N})^2} \,.
\label{eqn:transmission}
\end{equation}
The current across the junction at an applied voltage $V$ is consequently given as an integral over energy and momentum
\begin{equation}
    J_\sigma = \frac{2 e}{h} \int_{-\infty}^\infty \frac{dk_y}{2 \pi} \int_{-\infty}^\infty dE
    \big(f(E - eV) - f(E) \big) \, T_\sigma(E, k_y) \,,
\label{eqn:spin_current}
\end{equation}
with the Fermi-Dirac distribution $f(E) = 1 / (\e^{-\beta E} + 1)$.

The corresponding relations for the pocket around $M$ can be obtained by setting $k \rightarrow k + M$ and $\Delta \rightarrow - \Delta$.
Already from \cref{eqn:transmission} it is apparent that the $M$ pocket can not contribute to the transport across the \sAM-N junction: Modes with $k_y \sim M$ and $k_x > 0$ for $x < 0$ necessitate an imaginary $k_x$ for $x > 0$, i.e., are decaying with distance from the junction since $E, k_y$ are conserved and there are only electronic states with $k_y \sim \Gamma$ available for $x > 0$.
Thereby, the investiagted \sAM-N junction has similar properties as an FM-N junction and can be utilised to inject spins into a metal, while its zero net magnetization enhances the rigidity of the junction against stray fields.

\begin{SCfigure}
    \centering
    \includegraphics{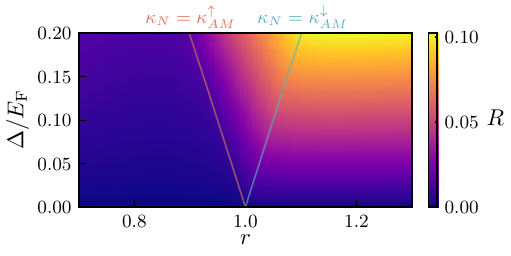}
    \hspace{2ex}
    \caption{Spin conversion factor as a function of Fermi surface mismatch and AM spin gap. We used \cref{eqn:spin_current} to calculate the spin current for a representative parameter set $\Ef^\text{N} = \Ef^\text{\sAM} = \Ef = 5$, $m_\text{\sAM} = 3$, $V = 0.2$, $Z = 0.4$ and variable $m_\text{N}$. The black dashed lines indicate the parameters, where the Fermi pockets partially coincide on the left and right hand of the junction.}
    \label{fig:spin_transport}
\end{SCfigure}

We calculate the spin dependent current at an inverse temperature $\beta = \infty$ and obtain a dependence of the spin conversion factor $R = \left| \frac{J^\uparrow - J^\downarrow}{J^\uparrow + J^\downarrow}\right|$ on the spin-valley polarization $\Delta$ and the Fermi surface mismatch $r = \kappa_{N}/\kappa_\text{\sAM}$ with $\kappa_\text{\sAM} = (\kappa^\uparrow_\text{\sAM} + \kappa^\downarrow_\text{\sAM}) / 2$ in \cref{fig:spin_transport}.
While the precise shape of the graph as well as the maximal values of the spin conversion factor are parameter dependent, we can classify its functional dependence on $\Delta$ and $r$ in three different regimes:
While the overall current increases with larger r since more propagating modes are provided on the right side of the junction, the effect on the two different spin species is almost equal as long as the Fermi surface radius of the semimetal $\kappa_\text{N} < \kappa_\text{\sAM}^\downarrow < \kappa_\text{\sAM}^\uparrow$.
When $\kappa_\text{\sAM}^\downarrow < \kappa_\text{N} < \kappa_\text{\sAM}^\uparrow$, the spin $\downarrow$ carriers cannot profit anymore from the additionally provided modes in the semimetal. While the $\downarrow$ current saturates at $\kappa_\text{N} = \kappa_\text{\sAM}^\downarrow$, the spin $\uparrow$ current continues to grow linearly until $\kappa_\text{N} = \kappa_\text{\sAM}^\uparrow$. 
Accordingly, the spin conversion factor reaches its maximum at exactly this value.
When increasing $\kappa_\text{N}$ even further we observe a gentle decay of $R$ as well as the overall current since the eigenspectra of the two parts of the junction get increasingly incompatible.
Hence, we propose a Fermi surface mismatch of $\kappa_\text{N} \gtrsim \kappa_\text{\sAM}^\uparrow$ for an optimal spin conversion.

The majority spin (in terms of transport across the junction) can be switched by contacting the \sAM with a semimetal with a circular pocket around $M$. Then the situation is exactly analogous to the case described in the previous paragraphs with $\uparrow \, \leftrightarrow \, \downarrow$ due to the \sAM spin symmetry.
The presence of two electron pockets in the N part of the junction leads to a competition between the spin transport via the different pockets: Since both pockets have the same transport characteristic and the pockets of the \sAM are related by symmetry, the net spin current across the junction is solely dependent on the difference in the filling and/or effective masses of the two pockets of the semimetal.
For a given parameter set, the total spin current can be obtained by subtracting the two appropriate phase space points in \cref{fig:spin_transport}.

\section{Constructing minimal models with \sAM groundstates}

\subsection{Design principles for minimal models}

We construct our Hamiltonian as follows: In sublattice space $\tau$ we want $\bvec Q$-separated and under sublattice exchange symmetry connected bands, i.e.
\begin{equation}
  \tau^x  H(\bvec k) {\tau^x}^\dagger = H(\bvec k + \bvec Q) \ .
  \label{eqn:symkkpQ}
\end{equation}
Also we want our Hamiltonian to be invariant under $C_4$ rotations, hence we set $Q=(\pi,\pi)$. Note that on the square lattice a chiral transformation already connects $H(\bvec k)$  and  $H(\bvec k + \bvec Q)$. Hence we formulate a toy model with two chiral partners of intra-sublattice hopping on the two sublattice sites:
\begin{equation}
  \text{Tr} \, \tau^z  H(\bvec k) = v_z (\bvec k) = - v_z (\bvec k + \bvec Q) \propto m_\sigma(\bvec k) \, ,
\end{equation}
where $m_\sigma$ is the form factor of the effective spin gap
and
\begin{equation}
  \text{Tr} \, \tau^0  H(\bvec k) = v_0 (\bvec k) =  v_0 (\bvec k + \bvec Q) \, .
\end{equation}
Note that $v_z(\bvec k)$ and $v_0 (\bvec k)$ can be traced back to the $A_1$ [(extended) $s$-wave] lattice representation:
\begin{align*}
  v_z &= \cos(k_x) + \cos(k_y) \, , \, \cos(k_x)\cos(2 k_y) + \cos(k_y)\cos(2 k_x) \, , \, + ... \\
  v_0 &= \text{const} \, , \, \cos(k_x)\cos(k_y) \, , \, \cos(2 k_x) + \cos(2 k_y) \, , \, + ...
\end{align*}
Furthermore we want the intra-sublattice hopping to be invariant under $C_4$ rotations
\begin{equation}
  |v_x(k_x, k_y)|^2 + |v_y(k_x, k_y)|^2  = |v_x(\pm k_x, (\pm) k_y)|^2 + |v_y(\pm k_x, (\pm) ky)|^2 \, .
\end{equation}
With \cref{eqn:symkkpQ} we find $- v_y(\bvec k) = v_y(\bvec k + \bvec Q)$.
This can be achieved by setting
\begin{equation}
  \text{Tr} \, \tau^x  H(k_x, k_y) =\text{Tr} \, \tau^y  H(k_x, -k_y)= v_x (k_x,k_y) =  v_y (k_x,k_y) \, .
\end{equation}
and 
\begin{equation}
v_y (\bvec  k) = \cos(k_x / 2 + k_y / 2) \ .
\end{equation}

\begin{figure}
    \centering
    \includegraphics[width=\columnwidth]{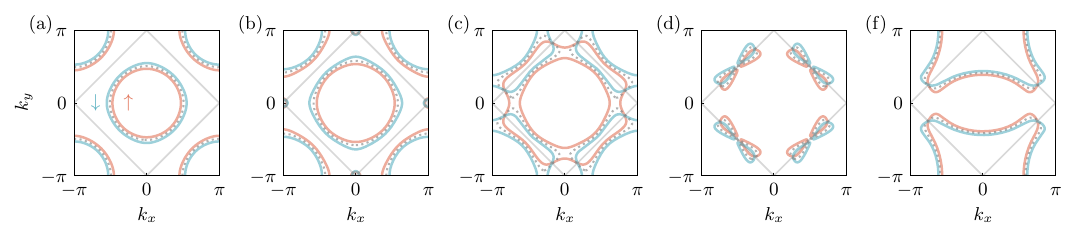}
    \caption{Fermi surface of the \sAM for the orbital model defined in \cref{eqn:hamiltonian_orbital} for different model parameter. The FS of the normal state is indicated by gray dotted lines.
    (a-d)~Spin split FS for $t_z = 1$, $t_x = t_x^\prime = 0.5$, $t_0 = 0$ and a magnetic order $\Delta = 0.3$ for $\Ef = [-2, -1.5, -1, -0.5]$.
    (f)~Same as~(a) but with $t_0 = 0.5$ breaking $C_4$ symmetry.
    In all cases, the momentum dependent spin splitting exhibits a symmetry protected node at $\cos(k_x) + \cos(k_y) = 0$ indicated by solid gray lines.}
    \label{fig:FS_orbital}
\end{figure}

\subsection{Node structure and extensions of the orbital model}
\label{SM:model_extension}

Like AMs with finite angular momentum ($p$-, $d$-, $f$-, $g$-wave), \sAMs likewise feature symmetry protected nodes in the magnetic order parameter to ensure full spin compensation. For the valley model discussed in the main text, the nodes are outside the validity range of the valley model and can hence be discarded.
For lattice models, where the \sAM symmetry is present at all energy scales, the momentum structure of the spin gap is fixed by the layer exchange symmetry and coincides with the lowest lattice harmonic transforming in the trivial $A_1$ irrep $\cos(k_x) + \cos(k_y)$.
This becomes apparent in \cref{fig:FS_orbital} upon varying the Fermi level of the system.

The node structure even persists the breaking of $C_4$ rotational symmetry by adding an additional term to the lattice Hamiltonian (cf.~\cref{latticemodel} in the main text):
\begin{equation}
\begin{split}
    H_\text{nematic} = & \, 2 t_0 \sum_{\bvec k \sigma}
    \crespinor_{\bvec k} \big[ \cos(k_x) - \cos(k_y) \big] \tau^0
    \annspinor_{\bvec k} \ .
\end{split}
\label{eqn:hamiltonian_orbital}
\end{equation}
Still momentum space translations by $M = (\pi,\pi)$ act as an orbital exchange symmetry, but now combined with a $C_4$ rotation, that is no longer part of the system's pointgroup $C_{2v}$. The corresponding spin symmetry reads $\mathcal S = \{C_4 \tau^x | C_2\}$.
This additional term leaves the nodes of the \sAM in \cref{fig:FS_orbital} untouched, since the $C_4$ symmetry is still part of the spin group symmetry.

\subsection{Flux lattice}
\label{SM:flux_model}
\begin{figure}
    \centering
    \includegraphics[width=\columnwidth]{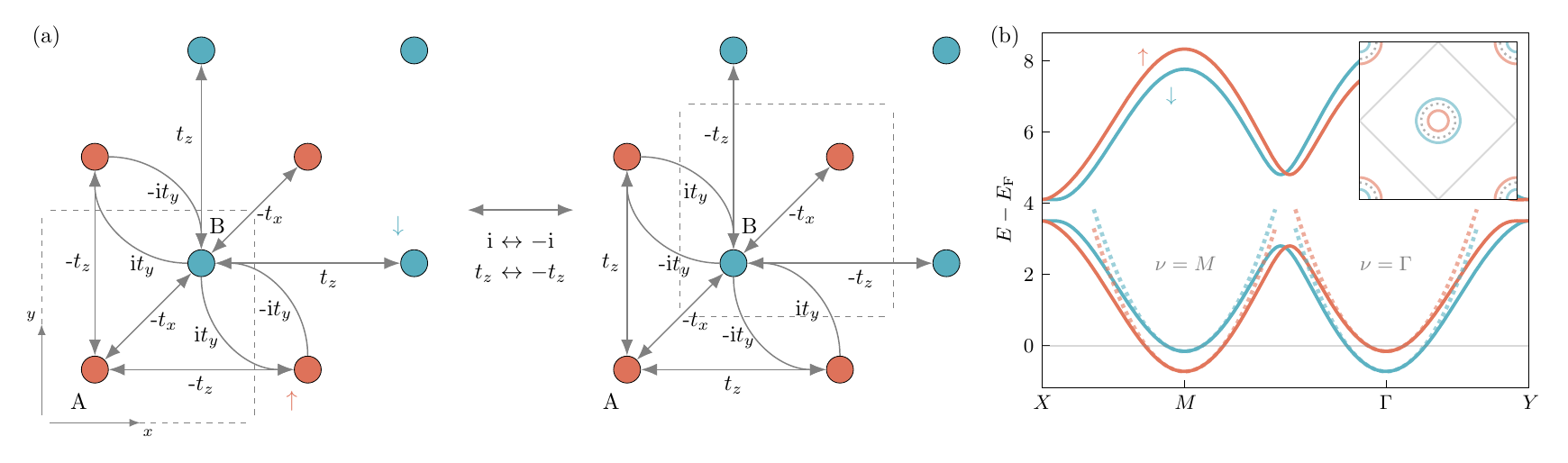}
    \caption{Lattice realization of the \sAM in a flux lattice. The hoppings on a square lattice leading to the Hamiltonian in \cref{eqn:flux_model} are visualized in~(a). The symmetry transformation under sublattice exchange $A \leftrightarrow B$ are indicated.
    The spin split bandstructure with a staggered spin polarization of the low energy effective valley model (indicated by parabolas) is depicted in~(b) for the parameter set $t_x = t_y = 0.5$, $t_z=1$, $\Ef=-3.8$ and a magnetic order defined in \cref{eqn:sAM_flux} with $\Delta = 0.3$. 
    The inset shows the Fermi surface of the normal state indicated by gray dashed lines, while the symmetry enforced nodal lines are given by solid lines. The spin up/down polarized Fermi sheets are depicted in red/blue.}
    \label{fig:flux_lattice}
\end{figure}

We exemplify the possibility of an $s$-wave altermagnet by considering a two orbital model on the square lattice. The hoppings in real space are visualized in \cref{fig:flux_lattice}.
The normal state Hamiltonian in momentum space reads
\begin{equation}
    H = \sum_{\bvec k \sigma}
    \begin{pmatrix} \ann_{\bvec k,A,\sigma} \\ \ann_{\bvec k,B,\sigma} \end{pmatrix}^\dagger
    H^\sigma(\bvec k)
    \begin{pmatrix} \ann_{\bvec k,A,\sigma} \\ \ann_{\bvec k,B,\sigma} \end{pmatrix} \ ,
\end{equation}
where the Hamiltonian in orbital basis is given by
\begin{equation}
    H^\uparrow(\bvec k) =
    \begin{pmatrix}
        - 2 t_z (\cos(k_x) + \cos(k_y) ) & - 2 t_x \cos(k_x / 2 + k_y / 2) - 2 t_y \I \cos(k_x / 2 - k_y / 2) \\
        - 2 t_x \cos(k_x / 2 + k_y / 2) + 2 t_y \I \cos(k_x / 2 - k_y / 2) & 2 t_1 (\cos(k_x) + \cos(k_y) )
    \end{pmatrix}\,,
\label{eqn:flux_model}
\end{equation}
and $H^\downarrow(\bvec k) = \mathcal{K} H^\uparrow(\bvec k)$ with complex conjugation $\mathcal{K}$. This ensures time reversal symmetry of the normal state Hamiltonian.
In the spinor space $\crespinor_{\bvec k} = ( \cre_{\bvec k,A,\uparrow} , \, \cre_{\bvec k, B, \uparrow} , \, \cre_{\bvec k, A, \downarrow} , \, \cre_{\bvec k, B, \downarrow})$ the Hamiltonian can be conveniently written in terms of Pauli matrices acting in orbital ($\tau^i$) and spin ($\sigma^i$) space 
\begin{equation}
    H= \sum_{\bvec k \sigma}
    \crespinor_{\bvec k} \left[ (
      v_z \tau^z
    + v_x \tau^x )  \sigma^0
    + v_y \tau^y  \sigma^z \right]
    \annspinor_{\bvec k} \,,
\label{eqn:hamiltonian}
\end{equation}
with the definitions
\begin{equation}
\begin{split}
    v_x &{}= - 2 t_x \cos(k_x / 2 + k_y / 2) \,, \\
    v_y &{}= 2 t_y \cos(k_x / 2 - k_y / 2) \,, \\
    v_z &{}= - 2 t_z (\cos(k_x) + \cos(k_y) ) \,.
\end{split}
\label{eqn:v_i}
\end{equation}
Inspecting the real space representation of this Hamiltonian (cf.~\cref{fig:flux_lattice}), we see that \cref{eqn:hamiltonian} corresponds to an extended $\pi$-flux square lattice.
To obtain the eigenspectrum of the Hamiltonian we can use the Pauli matrix algebra $\tau^i \tau^j = \delta_ij + \I \varepsilon_{ijk} \tau^k$ with the Kronecker $\delta$-symbol and the Levi-Civita-symbol $\varepsilon$. With this we can obtain the eigenvalues  of $H$ as the roots of the eigenvalues of $H^2$ to obtain $\epsilon_\sigma(\bvec k) = \pm \sqrt{|v_x|^2 + |v_y|^2 + |v_z|^2} - \Ef$. Since the Hamiltonian is block diagonal in spin space, we can easily convince ourselves, that it obeys Kramers degeneracy, i.e., all eigenstates feature a twofold spin degeneracy.

\subsubsection{Symmetry analysis}
Since symmetries are incremental for the characterization of magnetic states we (at least try to) list all symmetries of the Hamiltonian in \cref{eqn:hamiltonian} in the following.
Since the eigenspectrum is given by the norm of the $\Vec{v}$ vector, it is easy to show that all operations that act as a simple rotation in the $\Vec{v}$ space leave the spectrum invariant and are governed by the Pauli matrix rotation operator
\begin{equation}
    \mathcal R = \exp(\I \phi/2 \hat{n} \Vec{\tau}) = \tau^0 \cos(\phi / 2) + \I \hat{n} \Vec{\tau} \sin(\phi / 2) \ , 
\label{eqn:rotation}
\end{equation}
where $\hat{n}$ is a unit vector and indicates the axis of rotation in $\Vec{\tau}$ space.
However, these operations do not leave the Hamiltonian in \cref{eqn:hamiltonian} invariant and have to be supplemented with appropriate transformation behavior of the fermionic operators (like complex phases etc).
We further note, that hermiticity of the Hamiltonian dictates $v_i \in \mathbb{R} \, \forall i$.

\paragraph{Point group symmetries}
These symmetries are a direct consequence of the choice for $v_i$ in the given lattice model and are hence nonuniversal.
With \cref{eqn:v_i}, the Hamiltonian features $C_{2v}$ symmetry, i.e., a pointgroup comprised of in-plane inversion $C_2$ and the two diagonal mirror planes $S_{\langle 11 \rangle}$, $S_{\langle 1-1 \rangle}$.
These symmetries only act in momentum space and have a trivial representation both in spin and orbital space. It is crucial to note that there is no crystal symmetry that maps one orbital to the other. From this point of view, a fully compensated spin-split Fermi surface should not be possible in this model.

In the special case of $t_x = t_y$, the spectrum is additionally invariant under $C_4$ rotation and vertical and horizontal mirrors $S_{\langle 10 \rangle}$, $S_{\langle 01 \rangle}$. However, due to the magnetic flux through the lattice a $C_4$ rotation only leaves the Hamiltonian invariant if we additionally account for the complex phases of the AB hoppings. Since there is a gauge freedom in choosing the vector potential on the bonds to realize a $\pi$ flux through each plaquette, the concrete form this magnetic $\Tilde{C}_4$ symmetry is dependent on the employed gauge.
For the above choice in \cref{eqn:v_i}, $C_4$ simply rotates $\sigma^x \rightarrow - \sigma^y$ such that a following $C_4$ rotation in $\Vec{v}$ space $\Tilde{C}_4 = C_4 \otimes \I \tau^z  \sigma^0$ leaves the Hamiltonian invariant:
$\Tilde{C}_4 H \Tilde{C}_4^{-1} = H$.
\\
Again it is crucial that all spatial symmetries take a diagonal form in orbital space, i.e., only introduces orbital dependent phases and do not map one orbital to another.

\paragraph{Orbital symmetries}
Yet there are other symmetries in this system, that acquire an off-diagonal shape in orbital space. Since they act trivial in real space, they are irrespective of the actual choice of $v_i$ and are hence more general.
We proceed by setting $\Ef = 0$ since the filling does not change the central properties of the Hamiltonian but obscures some of the symmetries we want to investigate in this system.

\paragraph{Symmetries $v_x = v_y = 0$}
For this case the Hamiltonian is simply given by two copies of the square lattice Hubbard model with NN hopping of reversed sign.
This directly leads to the well known anti-unitary particle-hole (ph) symmetry, that is fulfilled for each sublattice separately
\begin{equation}
    \ph \alpha \crespinor_{i} \ph^{-1} = \alpha^* \exp(\I M \bvec r^\nu_i) \annspinor_{i}
\end{equation}
where $M = \pi / a (1, 1)^T$ and $\bvec r^o_i = a (i_x, i_y) + \bvec \delta^\nu$. $\bvec \delta^\nu$ is the sublattice position in the unit cell and $i_n \in \mathbb{Z}$ labels the respective unit cell in the square lattice.
Since we have two copies of the same system but inverted hopping signs, i.e., $\I \tau^y H -\I \tau^y = -H$, we additionally have the ph + sublattice exchange symmetry
\begin{equation}
    \ex \alpha \crespinor_{i} \ex^{-1} = \alpha^* \I \tau^y \annspinor_{i} \ .
\end{equation}
This acts local in momentum space and just flips the spectrum around the Fermi energy.
The combination of both symmetries gives $\sym = \ex  \ph$, that transforms as
\begin{equation}
    \sym \alpha \crespinor_{i} \sym^{-1} = \alpha \exp(\I M \mathbf r^o_i) \I \tau^y \crespinor_{i} \ .
\label{eqn:sym_flux}
\end{equation}
We note, that this operation does no longer exchange particles with holes.
It merely acts as a momentum space translation with $\bvec k \rightarrow \bvec k + M$ accompanied by a sublattice exchange.
Hence, it remains a symmetry in the non-half filled case at finite $\Ef$ and also for finite $v_x, v_y$. 
Pictorially half of the Brillouin zone (BZ) is connected to the other half shifted by $M$ via a sublattice exchange.

\paragraph{Symmetries for $v_z = 0$}
For $v_z = 0$ (and leaving aside the chemical potential) our system is bipartite and features the additional \textit{chiral symmetry}, that reads $\chiral = \tau^z  \sigma^0$ like it is well known for graphene and gives $\chiral H \chiral^{-1} = -H$.
We can convert this to an actual symmetry of the Hamiltonian, i.e., one that commutes with $H$, if we additionally consider the transformation of the fermionic operators
\begin{equation}
    \chiral \crespinor_i \chiral^{-1} = \tau^z \annspinor_i \ .
\end{equation}
We recognize, that this symmetry only rotates $\Vec{v}$ around the $z$-axis and hence does not involve a sublattice exchange.

\subsubsection{Impact on magnetic states}
Let us now consider an inter-sublattice AFM state, that introduces an additional term in the quasiparticle spectrum
\begin{equation}
    H_\text{AFM} = \sum_k \crespinor_{\bvec k} \Delta \tau^z  \sigma^z \annspinor_{\bvec k} \ .
\label{eqn:sAM_flux}
\end{equation}
Clearly this term breaks the usual time reversal symmetry (TRS) defined by the operator
\begin{equation}
    \TRS = \tau^0  \I \sigma^y \mathcal{K}
\end{equation}
since $\TRS (H + H_\text{AFM}) \TRS^{-1} = H - H_\text{AFM}$.
However, the combination $\sym \TRS$ remains a symmetry of the full Hamiltonian $\sym \TRS H_\text{AFM} (\sym \TRS)^{-1} = H_\text{AFM}$.
From this we obtain two important things:
(i)~The Hamiltonian is spin split since TRS is broken and we obtain spin polarized quasi-particle bands. This follows directly from the electronic eigenspectrum: Since also in the magnetic state the Hamiltonian remains block diagonal in spin space, we can write $\epsilon_\sigma(\bvec k) = \pm \sqrt{|v_x|^2 + |v_y|^2 + |v_z + \sigma \Delta|^2} - \Ef$.
(ii)~The magnetic state features perfectly compensated spin structure. This follows directly from
\begin{equation}
\begin{split}
    H \sym \TRS \ket{\bvec k \sigma} &{}= \sym \TRS H \ket{\bvec k \sigma} = \epsilon_\sigma(\bvec k) \sym \TRS \ket{\bvec k \sigma} = \epsilon_\sigma(\bvec k) \ket{M - \bvec k \overline \sigma} \,, \\
    H \sym \TRS \ket{\bvec k \sigma} &{}= H \ket{M - \bvec k \overline \sigma} =
    \epsilon_{\overline \sigma}(M - \bvec k)\ket{M - \bvec k \overline \sigma} \,,
\end{split}
\end{equation}
and is symmetry protected by $\sym$.
Hence, for every eigenstate at $\bvec k$ with spin $\uparrow$, there is a degenerate eigenstate with spin $\downarrow$ at $\bvec k + M$.
An exemplary spin-resolved Fermi surface is shown in \cref{fig:flux_lattice}, where the nodes in the spin gap along the $|\bvec k| = |M - \bvec k|$ line are indicated.

\paragraph{$v_z = 0$ case}
Like before, even though $\TRS$ is broken individually, the combination $\mathcal C \TRS$ is still a symmetry of the Hamiltonian and $(\mathcal C \TRS)^2 = -1$ ensures Kramer's degeneracy.
$\mathcal C$ acts trivial in momentum space, while $\TRS \bvec k \rightarrow - \bvec k$.
In the special case of \cref{eqn:v_i}, the Hamiltonian has inversion symmetry, which directly gives spin degenerate bands in the $v_z = 0$ case.
We can also understand this very intuitively from the vector in $\Vec{\tau}$ space:
For $v_z = 0$ the $z$ component of $\Vec{\tau}$ is exclusively given by $\pm \Delta$ for $\uparrow/\downarrow$. Consequently, $|\Vec{v}_\uparrow| = |\Vec{v}_\downarrow|$ and the system features spin degeneracy, since this equation is true $ \forall \, \bvec k$ separately.

\paragraph{Other possible magnetic states}
Besides a ferromagnet, there is no other possible magnetic state without translational symmetry breaking.
Especially, no conventional altermagnetic state can be found, since there is no lattice symmetry actually connecting the two sublattices.
This can be directly seen from the appearance of the operators in $\Vec{v}$ space:
All point group symmetries only rotate $\Vec{v}$ around the $z$ axis (i.e., are $\propto \tau^z$), so it is impossible to map $v_z \rightarrow - v_z$ by a pointgroup symmetry.

\subsection{\sAM with higher harmonics}
\label{SM:other_swave}

\begin{figure}
    \centering
    \includegraphics[width=\columnwidth]{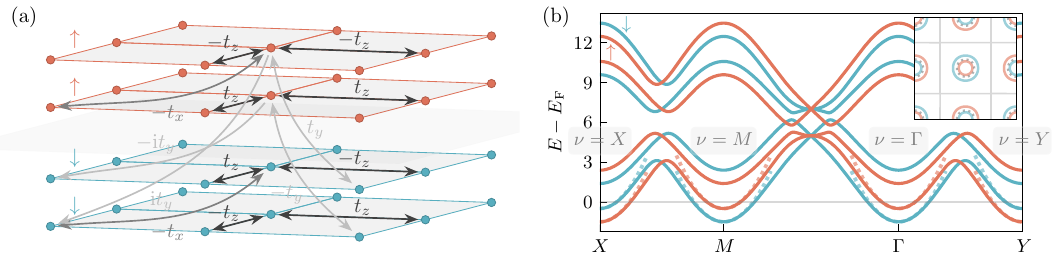}
    \caption{Four orbital model of an \sAM defined in \cref{eqn:ham_other_swave}. Panel~(a) presents a visualization of the hoppings, where each layer represents a different orbital and the \sAM state is given by an AFM order between the different layers as indicated by red/blue color.
    The correpsonding spin split bandstructure is shown in~(b) for $t_z = t_x = t_y = 1$, $\Ef = -6$ and the size of the spin polarization per layer is $\Delta = 1$. The low energy effective model consisting of 4 valleys at $\nu = X, \Gamma, M, Y$ is indicated by the dotted parabola for both spin species. The spin up/down polarized Fermi sheets are depicted in the inset in red/blue with symmetry enforced nodal lines of the spin gap indicated by solid gray lines. Unlike the two layer and flux model, the nodal lines are given by the next nearest neighbor rather than the nearest neighbor lattice harmonic.}
    \label{fig:GMXY_model}
\end{figure}

In addition to the spin gap structure presented in the main text, higher harmonics of the \sAM can be generated by constructing models with more complicated $\bvec k$ space symmetry. We consider a four orbital model and split the orbital Hilbert space in $2 \times 2$ blocks with the Pauli matrices $\tau^i (\nu^i)$ acting within (between) the blocks respectively
\begin{equation}
    \begin{split}
        H_0 = - \, & t_z \sum_{\langle ij \rangle_x, \sigma}
    \crespinor_{i, \sigma} \tau^z  \nu^0 \annspinor_{j, \sigma}
                        - \, t_z \sum_{\langle ij \rangle_y, \sigma}
    \crespinor_{i, \sigma} \tau^z  \nu^z \annspinor_{j, \sigma} \\
                - \, & t_x \sum_{\langle\langle ij \rangle \rangle, \sigma}
    \crespinor_{i, \sigma} \tau^x  \nu^0 \annspinor_{j, \sigma}
                + t_y \sum_{\langle\langle ij \rangle \rangle, \sigma}
    \crespinor_{i, \sigma} (\tau^z + \tau^y)  \nu^x \annspinor_{j, \sigma} \ .
    \end{split}
    \label{eqn:ham_other_swave}
\end{equation}
Here $\langle i j \rangle_{x/y}$ denotes the nearest neighbors in the $x/y$ direction and the spinor of fermionic creation operators $\crespinor_{i, \sigma} = \Big( \cre_{i, A, \sigma}, \, \cre_{i, B, \sigma}, \, \cre_{i, C, \sigma}, \, \cre_{i, D, \sigma} \Big)$ is a four dimensional vector defined in orbitals space carrying both unit cell index $i$ and spin $\sigma$ as external indices. A visualization of the hopping structure on the square lattice can be found in \cref{fig:GMXY_model}.
The kinetic Hamiltonian in momentum space representation reads
\begin{equation}
    H_0 = \sum_{\bvec k} \crespinor_{\bvec k, \sigma}
    \left[ (v_z^+ \tau^z + v_x \tau^x )  \nu^0
         + v_z^- \tau^z  \nu^z
         + v_y ( \tau^z +\tau^y)  \nu^x \right]
    \annspinor_{\bvec k, \sigma} \,,
\label{eqn:GMXY_model}
\end{equation}
with the definitions
\begin{equation}
\begin{split}
    v_x &{}= - 4 t_x \cos(k_x) \cos(k_y) \,, \\
    v_y &{}= 2 t_y \cos(k_x) \cos(k_y) \,, \\
    v_z^+ &{}= - 2 t_z \cos(k_x) \,, \\
    v_z^- &{}= - 2 t_z \cos(k_y) \,.
\end{split}
\end{equation}
The four dimensional spinor space representation of \cref{eqn:GMXY_model} takes the following form in momentum space:
\begin{equation}
    H_0 = \sum_{\bvec k, \sigma} \crespinor_{\bvec k, \sigma}
    {\small
    \begin{pmatrix}
        - 2 t_z \big(\cos(k_x) + \cos(k_y)\big) &
        - 4 t_x \cos(k_x) \cos(k_y) &
        2 t_y \cos(k_x) \cos(k_y) & - 2 t_y \I \cos(k_x) \cos(k_y)
        \\
        - 4 t_x \cos(k_x) \cos(k_y) &
        2 t_z \big(\cos(k_x) + \cos(k_y)\big) &
        2 t_y \I \cos(k_x) \cos(k_y) &
        - 2 t_y \cos(k_x) \cos(k_y)
        \\
        2 t_y \cos(k_x) \cos(k_y) &
        - 2 t_y \I \cos(k_x) \cos(k_y) &
        - 2 t_z \big(\cos(k_x) - \cos(k_y)\big) &
        - 4 t_x \cos(k_x) \cos(k_y)
        \\
        2 t_y \I \cos(k_x) \cos(k_y) &
        -2 t_y \cos(k_x) \cos(k_y) &
        - 4 t_x \cos(k_x) \cos(k_y) &
        2 t_z \big(\cos(k_x) - \cos(k_y)\big)
    \end{pmatrix}
    }
    \annspinor_{\bvec k, \sigma} \,.
\end{equation}

\subsubsection{Symmetry analysis}
The model enjoys $C_{4v}$ symmetry, where all pointgroup operations act trivially in orbital space.
However, inspecting momentum space translations we can see two different transformation behaviors in the orbital sector:
An $M$ point translation provides a symmetry of the Hamiltonian in combination with an exchange of the orbitals in the individual $2 \times 2$ block,
\begin{equation}
    \sym^{\vphantom 1}_M \alpha \crespinor_{\bvec k, \sigma} \sym_M^{-1} = \overline \alpha \, \tau^x  \nu^0 \crespinor_{\bvec k + M, \sigma} \,,
\end{equation}
while leaving the inter-block structure intact. This is in line with the previous symmetries observed in, e.g., \cref{SM:flux_model}.
An $X$ or $Y$ point shift, on the other hand, exchanges the individual blocks leading to a more complicated symmetry:
\begin{equation}
    \sym^{\vphantom 1}_Y \alpha \crespinor_{\bvec k, \sigma} \sym_Y^{-1} = \overline \alpha \, \tau^z  \nu^x \crespinor_{\bvec k + Y, \sigma} \,,
\end{equation}
and an analogous formula can be found for $X$ as demanded by $C_4$ symmetry:
\begin{equation}
    \sym^{\vphantom 1}_X \alpha \crespinor_{\bvec k, \sigma} \sym_X^{-1} = \overline \alpha \, \tau^y  \nu^x \crespinor_{\bvec k + X, \sigma} \,.
\end{equation}
This symmetry exchanges both intra- and inter-block orbitals.

\subsubsection{Altermagnetic states}
\begin{figure}

\begin{tikzpicture}

\def\arraystretch{2.} 
\setlength{\tabcolsep}{1.2em} 
\matrix[ampersand replacement=\&]{
\node[anchor=south] (species) at (0, 1.1) [shape=rectangle]{
\begin{tabular}{|c | c c c |}
 \hline
  & $\sym_X$ & $\sym_Y$ & $\sym_M$ \\ [0.5ex] 
 \hline
 $\Delta_0$ & $-$ & $-$ & $+$ \\ 
 $\Delta_1$ & $+$ & $-$ & $-$ \\
 $\Delta_2$ & $-$ & $+$ & $-$ \\ [1ex] 
 \hline
 \end{tabular}};
\node[anchor=north west] at ($(species.north west)-(.7,-.3)$) {(a)};
\& 
\node[anchor=south] (pdf) at (0,0) {\includegraphics{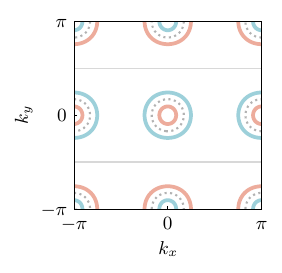}};
\node[anchor=north west] at (pdf.north west) {(b)};
\&
\node[anchor=south] (pdf) at (0,0) {\includegraphics{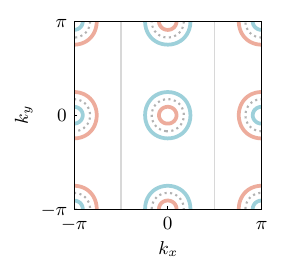}};
\node[anchor=north west] at (pdf.north west) {(c)}; \\
};
\end{tikzpicture}
\caption{Different compensated magnetic states of the 4 orbital model defined in \cref{eqn:ham_other_swave}.
(a)~Eigenvalues of the different AM states in \cref{eqn:AM_states} under the symmetry operations building the spin group of the \sAM.
(b,c)~Spin split Fermi surface with a magnetic order transforming according to $\Delta_1$~(b) or $\Delta_2$~(c). Both magnetic states break $C_4$ symmetry, which is directly reflected in \sAM nodal line structure (solid gray lines. The model parameters are the same as for \cref{fig:GMXY_model}.}
\label{tab:trafo_behavior}
\end{figure}

Due to the symmetry restrictions of the model, one can think of three different compensated magnetic structures on the lattice:
\begin{equation}
\begin{split}
    \Delta_0 &{}\propto \sum_{\bvec k, \sigma} \sigma^z_{\sigma\sigma} \, \crespinor_{\bvec k, \sigma} \tau^0  \nu^z \annspinor_{\bvec k, \sigma} \,,\\
    \Delta_1 &{}\propto \sum_{\bvec k, \sigma} \sigma^z_{\sigma\sigma} \, \crespinor_{\bvec k, \sigma} \tau^z  \nu^0 \annspinor_{\bvec k, \sigma} \,,\\
    \Delta_2 &{}\propto \sum_{\bvec k, \sigma} \sigma^z_{\sigma\sigma} \, \crespinor_{\bvec k, \sigma} \tau^z  \nu^z \annspinor_{\bvec k, \sigma} \,.
\end{split}
\label{eqn:AM_states}
\end{equation}
While all of these states are spin compensated by symmetry, the different eigenvalues of $\Delta_{1,2}$ with respect to $\sym_{X, Y}$ indicate a $C_4$ symmetry breaking for these states (cf.~\cref{tab:trafo_behavior}).
Consequently, $\Delta_0$ is the only AM state conserving $C_4$ symmetry and the nodes of the spin gap are determined by the transformation behavior of the order parameter shown in \cref{tab:trafo_behavior}.
The resulting spin split Fermi surface is notably distinct from the one depicted in \cref{fig:FS_orbital}:
While both transform trivially under all symmetry operations of $C_{4v}$ and are thus classified by the same irrep $A_1$, the nodal structure is given by two different lattice harmonics of the underlying square lattice structure:
Due to the orbital polarization of the electronic eigenstates, $\Delta_0(\bvec k) \propto \cos(k_x) \cos(k_y)$ (corresponding to a next nearest neighbor formfactor) compared to $\Delta(\bvec k) \propto \cos(k_x) + \cos(k_y)$ (corresponding to a nearest neighbor formfactor) for the model in \cref{SM:flux_model} and the main text.

Unlike the SC analog, the different harmonics are not allowed to mix for the \sAM to preserve spin compensation and the nodal lines are symmetry enforced by the delicate momentum space translation symmetry in the systems.

\subsection{Extension to 1D systems}
\label{SM:1D}
\begin{figure}
    \centering
    \includegraphics[width=\columnwidth]{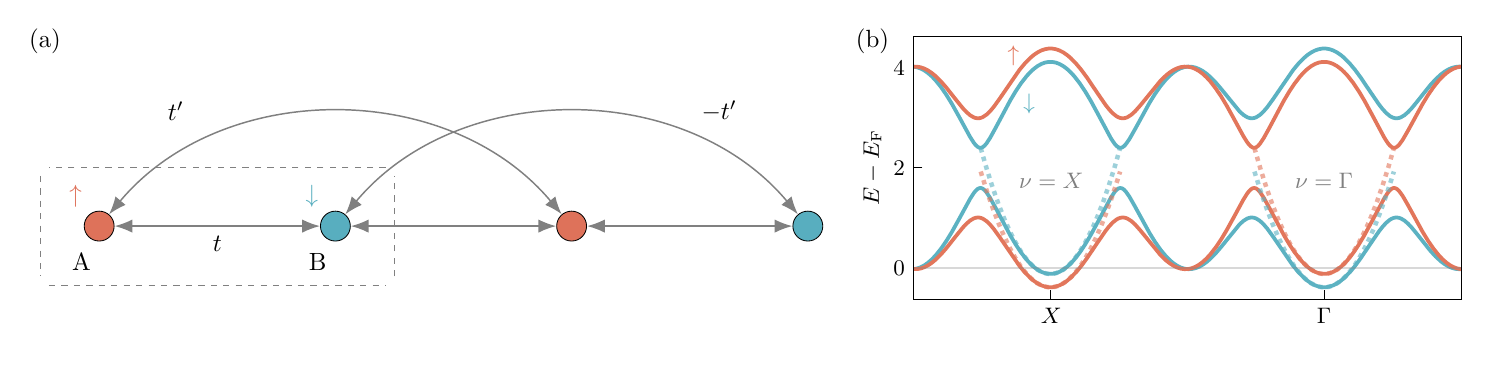}
    \caption{Model of an \sAM in one spatial dimension as defined in \cref{eqn:model_1D} with a visualization of the real space hopping structure~(a) and spin split bandstructure~(b) for $t' = 0.5 t$, $\Ef = -2t$ and a magnetic order $H_\text{\sAM} = \sum_{ k, \sigma} \crespinor_{ k, \sigma} \Delta \tau^z \sigma^z_{\sigma\sigma} \annspinor_{ k, \sigma}$ with $\Delta = 0.2$.}
    \label{fig:1D_model}
\end{figure}

Since the spin compensation does not rely on a point group symmetry (like rotations), the concept of $s$-wave altermagnetism can be easily extended to 1D systems, where the dimensional reduction excludes the existence of conventional collinear altermagnetic states~\cite{Roig2024m}.
We consider electrons confined to 1D with kinematics given by the normal state Hamiltonian
\begin{equation}
    H_0 = \sum_{i, \sigma} \left[ - t \, \cre_{2i, \sigma} \ann_{2i+1, \sigma}
                                     - t'  \, \cre_{2i, \sigma} \ann_{2i+2, \sigma}
                                     + t' \, \cre_{2i+1, \sigma} \ann_{2i+3, \sigma}\right] + \text{h.c.} ~,
\label{eqn:model_1D}
\end{equation}
with a (next) nearest neighbor hopping $t$ ($t'$). The momentum space Hamiltonian in the two site unit cell reads
\begin{equation}
    H_0 = \sum_{ k, \sigma} \crespinor_{ k, \sigma} [ -2 t \tau^x \cos(2k) - 2 t' \tau^z \cos(k)] \annspinor_{ k, \sigma} \,,
\end{equation}
with the spinor $\crespinor_{k s} = \big(\cre_{k, A, \sigma}, \, \cre_{k, B, \sigma} \big)$ and is invariant under the transformation
\begin{equation}
    \sym \alpha \crespinor_{k, \sigma} \sym^{-1} = \alpha \, \tau^x \crespinor_{k +  X, \sigma} \ . 
\end{equation}
A staggered magnetization $\Delta \propto \sum_{k, \sigma}\sigma^z_{\sigma\sigma} \crespinor_{k, \sigma} \tau^z \annspinor_{k, \sigma} $ on the two sites generates the spin split bandstructure depicted in \cref{fig:1D_model} with symmetry protected bandcrossings at $k = \pm \, X/2$ and full spin compensation.
This is to the best of our knowledge the first instance of non-relativistic momentum dependent spin splitting in 1D.

\section{Valley model for \sAM on the hexagonal lattice}
\label{SM:hexagon_valley}
\begin{SCfigure}[3.8][h]
    \includegraphics{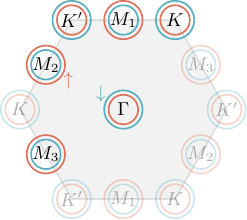}
    \hspace{2ex}
    \caption{Valley model for \sAM on the hexagonal lattice. We do not specify a symmetry connecting the $M$ ($\color{bdiv2-7}\uparrow$) to the $K$ and $\Gamma$ ($\color{bdiv2-2}\downarrow$) valleys here, but only sketch the spin polarization required for an extended $s$-wave spin gap.}
    \label{fig:hexagonal-sam}
\end{SCfigure}
\Cref{fig:hexagonal-sam} presents a possible \sAM state on the hexagonal lattice. The three valleys $M_{1,2,3}$ attain spin-up polarization, and the three valleys $K,K',\Gamma$ spin-down.

\end{document}